\documentclass[sigconf]{acmart}

\usepackage{array}
\usepackage{booktabs}
\usepackage{placeins}
\usepackage{xcolor}
\setcopyright{none}
\usepackage{listings}
\usepackage{soul}
\usepackage{enumitem}
\usepackage{tikz}
\usetikzlibrary{arrows.meta}

\lstdefinestyle{codingprompt}{
  basicstyle=\ttfamily\tiny,
  breaklines=true,
  columns=fullflexible,
  keepspaces=true,
  showstringspaces=false,
  frame=single,
  xleftmargin=0.5em,
  xrightmargin=0.5em
}

\providecommand{\revisioncolor}{black}

\AtBeginDocument{%
  \DeclareCaptionFont{revisioncolor}{\color{\revisioncolor}}%
  \captionsetup{font=revisioncolor}%
}

\begin{document}

\title[Value-Sensitive Delegation in Everyday AI Agent Use]{Value-Sensitive Delegation in Everyday AI Agent Use: Evidence from OpenClaw}

\author{Renkai Ma}
\authornote{Both authors contributed equally to this research.}
\email{mark@ucmail.uc.edu}
\affiliation{%
  \department{School of Information Technology}
  \institution{University of Cincinnati}
  \city{Cincinnati}
  \state{Ohio}
  \country{United States}
}

\author{Ruyuan Wan}
\authornotemark[1]
\affiliation{%
  \department{College of Information Sciences and Technology}
  \institution{The Pennsylvania State University}
  \city{University Park}
  \state{Pennsylvania}
  \country{United States}
}

\author{Xuan Lu}
\affiliation{%
  \department{College of Information Science}
  \institution{University of Arizona}
  \city{Tucson}
  \state{Arizona}
  \country{United States}
}

\author{Fan Yang}
\affiliation{%
  \institution{University of South Carolina}
  \city{Columbia}
  \state{South Carolina}
  \country{United States}
}

\author{Chen Chen}
\email{chechen@fiu.edu}
\affiliation{%
  \department{Knight Foundation School of Computing and Information Sciences}
  \institution{Florida International University}
  \city{Miami}
  \state{Florida}
  \country{United States}
}

\author{Lingyao Li}
\email{lingyaoli@arizona.edu}
\affiliation{%
  \department{College of Information Science}
  \institution{University of Arizona}
  \city{Tucson}
  \state{Arizona}
  \country{United States}
}

\renewcommand{\shortauthors}{Ma et al.}

\begin{abstract}
Users increasingly delegate work to autonomous AI agents, yet evaluations typically measure task completion rather than the values users prioritize. Using Value Sensitive Design, we analyzed, with LLM assistance, 73,093 first-person Reddit posts about using OpenClaw, each for its human value, agent aspect, value fulfillment, and user outcome. The 21 values form six value groups, including Autonomous, Dependable, and Affordable Operation, Bounded Reach, Reviewability, and Equitable Access. Relative to each aspect's corpus share, values clustered not at the agent's outputs but at the operating conditions users set around a run. Values were usually met where users described what the agent delivered, in five of six groups, and mostly unmet where users described supervising it, in all six groups. We conceptualize this pattern as \emph{value-sensitive delegation}. Supporting human values requires attention not only to what an agent accomplishes, but to the conditions users set around delegation, including cost, access, and oversight.
\end{abstract}





\maketitle

\section{Introduction}

An autonomous agent is an AI system that independently executes tasks and manages workflows on behalf of a user. Users increasingly delegate real work to these agents\footnote{We use ``AI agents'' throughout to refer to autonomous, large language model (LLM)-based agents. Prior work uses several labels for the same class of system, including ``GUI agent,'' ``LLM-based agent,'' and ``multi-agent system''~\cite{chen2025toward, yehudai2025survey, naik2025earlyadopters}, as do the posts we analyze, and we preserve the posts' wording when quoting them.}. OpenClaw~\cite{cnbc2026openclaw},\footnote{\url{https://github.com/openclaw/openclaw}} an open-source AI agent that runs on a user's own computer and performs tasks through chat apps such as WhatsApp, exemplifies this class of system. Released in November 2025, it had drawn more than 370,000 GitHub stars by August 2026~\cite{openclaw2026repo}. Such an agent differs from a typical conversational assistant in where execution occurs and when a user can review it~\cite{he2025plan, zhou2026checking}. A typical conversational assistant asked to fix a bug returns a patch for the user to inspect, whereas an agent reads the repository, edits the files, runs the tests, and commits the result under the user's credentials, with few checkpoints in between~\cite{cheng2026mapping, naik2025earlyadopters}. We call one such stretch of delegated work, from the user's instruction to the outcome the user reads afterward, a \emph{run}. Consequently, a chatbot that misreads a request returns a poor answer, whereas an agent that misreads one can read files the user never exposed~\cite{su2025survey}, exhaust an allowance the user cannot replenish, or delete work the user cannot recover~\cite{zhang2026consequences}. An agent can do all three even though human--AI interaction design guidelines have long asked that AI systems make their limits legible and support recovery when they fail~\cite{amershi2019guidelines}.

How AI agents are evaluated follows how they are built. Technical research decomposes them into modules that separate memory from planning, and both from the actions that call external tools~\cite{wang2024survey, xi2025rise}. Agent evaluations follow this decomposition, scoring whether a workflow completes a task~\cite{mohammadi2025evaluation, liu2024agentbench, zhou2024webarena, xie2024osworld, jimenez2024swebench, yao2025taubench}, while human--AI alignment often scores moral judgments on prompts~\cite{hendrycks2021ethics}. However, a benchmark score reflects the construct its evaluation was built to measure~\cite{raji2021everything, hutchinson2022evaluationgaps}, and few evaluations measure users' experience of these modules. HCI research has begun to examine such experience, showing that how and when a user is involved changes what the user and the agent achieve together~\cite{he2025plan, kim2026interventions} and how far the user overrelies on the agent~\cite{he2025plan, bucinca2021trust}. Less is known about how users meet these modules together, in their own everyday work. 

We investigate this experience through Friedman et al.'s \textit{Value Sensitive Design} (VSD), which holds that design should account for human values throughout development and defines a value as what a person or group considers important in life~\cite{Friedman1996Value-sensitiveDesign, friedman2006vsd}. VSD argues that the properties designers build into a technology more readily support some values and hinder others, yet whether a value is realized depends on the goals of the people interacting with it~\cite{friedman2006vsd}. An autonomous agent changes where that interaction happens, because it acts between the user's instruction and the outcome the user receives, so a value can be realized while the user is away. We therefore adopt VSD as our analytical lens. We ask which values users invoke when an agent acts for them, and which agent aspects\footnote{We use ``agent aspect'' to refer to the part of an agent system a user is dealing with when a value comes into play, which may be the model that produces an output, the account that meters a run, or the credentials that let an agent reach a file.} those values attach to.

Prior work applies VSD to AI primarily through design guidelines~\cite{sadek2024guidelines, umbrello2021mapping, cociancig2026toward} and interviews with AI practitioners~\cite{sadek2025challenges}, while separate studies of responsible-AI practice examine how organizations translate values into work~\cite{wang2023designingrai, varanasi2023hodgepodge}. That evidence comes from the developers who build AI systems, not from the users who delegate work to them, and as agents take on longer chains of action with fewer checkpoints, who answers for what an agent does becomes harder to settle from the developer's side alone. We thus ask two research questions:

\begin{itemize}[leftmargin=*, nosep]
    \item \textbf{RQ1.} \textit{What human values surface in users' first-person experience of using an autonomous AI agent like OpenClaw and which agent aspects do those values attach to?}
    \item \textbf{RQ2.} \textit{What outcomes do users attribute to agent use when those values are met and when they are not?}
\end{itemize}

To answer these questions, we collected Reddit posts about OpenClaw, screened them for first-person accounts, and coded each for its human value, agent aspect, value fulfillment, and user outcome. The schema's 21 values comprise 12 of the 13 values in Friedman et al.'s list~\cite{friedman2006vsd} and nine study-specific values, such as affordability and meaningful human control, that we developed from the corpus for delegating work to an agent~\cite{ledantec2009values}. We mapped all 21 values onto six value groups. RQ1 analyzes the 73,093 coded posts and RQ2 the 44,767 of them that carry an attributable user outcome. Five of the 21 values accounted for 70.8\% of posts. The agent aspects most overrepresented relative to their corpus share were the operating conditions users set around a run, namely what it cost, what it could reach, what it recorded, when it had to ask, and how it was installed. The model core, the underlying LLM that produces the agent's output, was not among them (Section~\ref{sec:rq1_aspects}). Fulfillment varied from 77.3\% met in Autonomous Operation to 42.8\% in Equitable Access, and resource accounting had the lowest met rate of all 18 agent aspects at 35.3\% (Section~\ref{sec:rq1_fulfillment}). Value fulfillment split what an agent delivered from what delegating to it cost and risked, with the post's value met in 67.7\% of task-effectiveness posts but in 34.7\% of resource-burden and 10.7\% of risk-exposure posts (Section~\ref{sec:rq2_fulfillment}). Value fulfillment and violation were therefore registered in these posts at agent aspects that the evaluations we reviewed do not cover~\cite{raji2021everything, hutchinson2022evaluationgaps}, and posts about what an agent could do on its own carried different fulfillment rates from posts about whether it could be relied on, a separation prior work on trust and reliance draws but agent evaluation does not~\cite{bucinca2021trust, vereschak2024trust}. Because these values sat in what users had set before a run rather than in what the run returned, we discuss them as \emph{value-sensitive delegation}, a relationship between a human value and the condition that carries it.

Our contributions are threefold: (1) An empirical characterization of the human values that surface in 73,093 first-person posts of OpenClaw use, and of where in an agent each of those human values attaches. (2) The conceptualization of \emph{value-sensitive delegation}, showing that values fell short at the operating conditions governing cost, reach, and setup at higher rates than at the model core of the OpenClaw agent, and that posts about an agent's reach, its reliability, its costs, and what it delivered carried systematically different fulfillment rates. That difference shifts the target of value-alignment work from the model to the boundaries users set around it. (3) Design implications keyed to the agent aspects where the shortfalls concentrate.


\section{Related Work}
We situate this study in two bodies of work. Section~\ref{sec:rw_agents} reviews how autonomous agents are evaluated and where such evaluations stop short of user experience. Section~\ref{sec:rw_vsd} reviews how VSD has been applied to AI and why that work has not reached the experience of using an agent.

\subsection{Evaluating Autonomous AI Agents}
\label{sec:rw_agents}
Using an autonomous agent is an act of \textit{delegation}, where a user hands a task to a system, decides how much autonomy to grant, and rejoins the work where they reserved a say. Early work on software agents and automation defined delegation as relying on another's action to reach a goal and distinguished levels of delegation by how much of the task the delegate decides~\cite{castelfranchi1998delegation}. Human-factors research described the handover along two axes, which functions a machine takes over and how far it automates each of them~\cite{parasuraman2000model}, and proposed delegation interfaces for supervisors to set the scope of automated systems~\cite{miller2007delegation}. More recent work asked which tasks people are willing to delegate to AI and proposed that motivation, difficulty, risk, and trust determine that willingness~\cite{lubars2019delegability}. LLM-based agents extend this handover, because a single instruction can start a run that edits files, spends money, and contacts external services before the user sees the results~\cite{chan2023harms}. Yet, the evaluation of these agents still centers on the technical question of whether an agent can complete a task in a given test environment. For example, benchmarks like AgentBench~\cite{liu2024agentbench}, WebArena~\cite{zhou2024webarena}, and OSWorld~\cite{xie2024osworld} test agents across interactive environments, websites, and desktop applications. Others apply this execution-based logic to domain-specific behavior, such as resolving GitHub issues (e.g., SWE-bench~\cite{jimenez2024swebench}) or reaching a correct database state during customer service interactions (e.g., $\tau$-bench~\cite{yao2025taubench}). While observable execution results verify task completion and enable technical capability comparisons, benchmark scholarship cautions against treating specific task performance as evidence of broader capability when relevant contexts are omitted~\cite{raji2021everything,hutchinson2022evaluationgaps}. For AI agents, this omitted context is the user's experience. A run scored as complete can still have taken supervision the score does not record, hidden further agent actions from the user, and left effects the user had to undo.

Security research evaluates these agents by architectural layer to identify where failures originate. For example, an analysis of 470 OpenClaw advisories found the dominant weakness was per-layer trust enforcement rather than a unified policy boundary~\cite{suwansathit2026security}. Other work has separated cognitive, execution, and information-system risks~\cite{ying2026uncovering}, traced attacks that propagate from prompt processing through tool invocation~\cite{wang2026assistant}, and demonstrated that poisoning one dimension of an OpenClaw deployment's persistent state increases attack success~\cite{wang2026your}. While this work shows that failures concentrate in identifiable parts of a system, it identifies them through advisory analysis and adversarial probing and scores them as attack success or exploitability rather than as issues end users encounter during daily use.

Complementing these technical evaluations, HCI evaluates autonomous AI agents as interactive systems. Human--AI interaction design guidelines set out what such systems owe a user, including legible capabilities and support for error recovery~\cite{amershi2019guidelines}, protection against overreliance~\cite{bucinca2021trust}, and trust built through human actors rather than system features alone~\cite{vereschak2024trust}. Research has begun mapping these requirements onto human--agent interaction by establishing design spaces for computer-use agents~\cite{cheng2026mapping}, studying early adopters balancing autonomy with human oversight~\cite{naik2025earlyadopters}, and analyzing the failures and workarounds of GUI web-browsing agents~\cite{zhang2026consequences}. However, to our knowledge, no shared framework has yet emerged for comparing how specific aspects of an AI agent shape user-reported experiences and outcomes. Our study addresses this gap by analyzing first-person posts of everyday OpenClaw use on Reddit.

\subsection{VSD and the Lived Experience of Agent Use}
\label{sec:rw_vsd}
Autonomous AI agents do more than complete tasks. Because a run reaches a user's files, credentials, and accounts, using an agent raises the question of whether end users retain meaningful control over delegated work, and a completion score does not answer it. VSD was built for questions of that kind, and pursues them through conceptual, empirical, and technical investigations that run throughout design~\cite{friedman2006vsd}. It asks what a technology supports or undermines for its stakeholders, not only whether the technology works.

Applying VSD to AI development remains difficult. Studies of responsible-AI practice show that unclear roles and organizational structures impede translating human values into practice~\cite{wang2023designingrai,varanasi2023hodgepodge}, while responsible-AI toolkits incorporate values through collaborative and educational features~\cite{sadek2024guidelines}. Responsible-AI and participatory-AI research continues to ask whose values a system represents~\cite{gabriel2020artificial, jakesch2022different, sadek2025challenges} and how stakeholders should influence design~\cite{birhane2022participatory, sloane2022participation, delgado2023participatory, kirk2024prism}. Because this evidence comes from design processes, practitioner intentions, and values elicited outside everyday use, it does not reveal which values end users experience as supported or undermined after an agent acts.

Work extending VSD helps distinguish design intentions from end-user experience, recognizing that predefined value classifications can obscure local expressions of value, whereas engagement with lived experience supports value discovery~\cite{ledantec2009values}. Computational research offers a complementary route by evaluating value-relevant model behavior, encoding moral judgments about text scenarios~\cite{hendrycks2021ethics}, testing safety consistency~\cite{wang2024fakealignment}, or linking diverse participant profiles to live LLM conversations~\cite{kirk2024prism}. However, its units of analysis remain elicited scenarios, test responses, and model conversations rather than end users' accounts of agents in everyday settings. Neither route yields an empirical account linking a concrete agent aspect to the value a user experienced, to whether that value was supported or undermined, and to the outcome reported in naturally occurring use. Our study provides that account.

\section{Methods}
\label{sec:methods}
To answer our research questions, we conducted a value-centered analysis of first-person Reddit posts about the use of OpenClaw. This section describes our data preparation (Section~\ref{sec:data_collection}), coding schema development and application (Section~\ref{sec:data_analysis}), human validation (Section~\ref{sec:human_validation}), statistical and qualitative analysis (Section~\ref{sec:statistical_analysis}), and ethics (Section~\ref{sec:ethics_availability}).

\subsection{Data Preparation}
\label{sec:data_collection}
We collected Reddit posts and comments\footnote{For simplicity, the rest of this paper uses ``posts'' to refer to both Reddit posts and comments.} created between January 31, 2026 and April 26, 2026. We set the start date at the cutoff for use of the OpenClaw name and the end date at our final data retrieval. We first collected candidate posts and comments with Brandwatch\footnote{\url{https://www.brandwatch.com}}, a social-media listening platform whose full-archive search returned Reddit IDs. We then used the Reddit PRAW API\footnote{\url{https://praw.readthedocs.io}} to retrieve the full content of each matched post and comment. The search used the case-insensitive keywords ``openclaw,'' ``clawdbot,'' and ``moltbot,'' the agent's current and earlier names, yielding a corpus of 1,100,308 posts, comprising 38,753 original posts and 1,061,555 comments.

\subsection{Coding Schema Development \& Application: Agent Aspect, Human Value (Group), and User Outcome}
\label{sec:data_analysis}
\textbf{Stage~1: Relevance screening for first-person experience.}
In Stage~1, we applied a screening-only prompt that classified each post as a first-person experience of using or attempting to use OpenClaw, a secondhand observation, or too thin to judge, keeping only the first-person posts (Appendix~\ref{app:prompt-design} defines the three categories). Stage~1 retained 187,479 posts, and Section~\ref{sec:human_validation} reports its validation.

\textbf{Stage~2: Value-centered coding.} We call this stage value-centered because it anchored the coding on the human value. We first identified the primary human value at stake in an excerpt, then mapped the mentioned agent aspect and any resulting user outcome to that same excerpt. Every label was grounded in a single excerpt. Value fulfillment was coded as met when the excerpt described the value as supported, and not met when it described the value as undermined, threatened, or unavailable. A post was retained only if it passed both stages and its three required fields, the human value, the agent aspect, and the value fulfillment status, were all valid; the user outcome field was optional. Appendix~\ref{app:prompt-design} summarizes the coding instructions. We grounded the schema's three coding dimensions in existing literature to give a standardized vocabulary across a large dataset.

\textbf{(1) Agent aspects.} We defined 18 agent-aspect categories (Table~\ref{tab:groups} and Appendix~\ref{app:prompt-design}) identifying the system component or configurable boundary implicated in the value-centered excerpt. The categories drew on prior technical literature, such as decompositions of agent architecture~\cite{wang2024survey,xi2025rise,sumers2023cognitive} and automation~\cite{parasuraman2000model}, along with HCI guidance on human oversight and recovery~\cite{amershi2019guidelines}, a GUI-agent evaluation framework~\cite{chen2025toward}, and AI agent security risks~\cite{su2025survey}. We refined these definitions during codebook development to keep agent aspects distinct from human values and user outcomes.

\textbf{(2) Human values.} We developed the value taxonomy through a deductive--inductive process~\cite{hsieh2005three}. We used 12 of the 13 values in Friedman et al.'s list~\cite{friedman2006vsd} as sensitizing concepts, omitting \textit{courtesy} because it concerns interpersonal politeness rather than how users delegate work to an agent. Inductively, we read a separate 50-post sample to identify human--agent interaction concerns absent from the deductive list~\cite{ledantec2009values}. We consolidated these into nine additional values, including meaningful human control, dependability, and contextual integrity, drawing from prior work~\cite{santonidesio2018meaningful,nissenbaum2004privacy,hummel2021data,avizienis2004basic}. Coder review of a 400-post sample (Section~\ref{sec:human_validation}) clarified definitions before full coding, and those review discussions acted as the kind of practice that surfaces values inside a team~\cite{shilton2013valueslevers}. Appendix~\ref{app:schema} defines all 21 values.

\textbf{(3) User outcomes.} We defined nine user outcomes (Table~\ref{tab:groups}) separately from the human values and agent aspects. We drew initial constructs from prior work on usability~\cite{iso2018usability}, user experience~\cite{hassenzahl2006ux}, workload~\cite{hart1988nasa}, trust~\cite{lee2004trust}, and automation~\cite{parasuraman1997humans}. We then refined the boundaries between those constructs using user examples from the 400-post codebook-development sample, so that the categories captured the consequences users actually attributed to agent use. A value captured what mattered in the excerpt, whereas an outcome recorded the consequence the author reported.

Using a large language model to apply a codebook at this scale trades coder time against a labeling error the researchers do not observe directly, so the procedure needs its own validation~\cite{schroeder2025qualitative}. We applied GPT-5 mini with the Stage~2 prompt (Appendix~\ref{app:prompt-design}) to the retained posts, keeping 73,797 of them (Section~\ref{sec:human_validation} reports agreement with human coders). RQ1 analyzed the 73,093 of those posts assigned to one of the 21 predefined values; we excluded posts with open-coded values because they lacked a documented mapping to the taxonomy. RQ2 analyzed the 44,767 posts within that RQ1 sample that carried an attributable user outcome. A missing outcome indicates no attributable consequence was coded, not a neutral outcome. Appendix~\ref{app:corpus-construction} summarizes sample construction.

For group-level analysis, we used an affinity diagram to organize the 21 value definitions into six value groups, each covering a distinct facet of delegated use (Table~\ref{tab:groups}). For example, Autonomous Operation clusters autonomy, human welfare, and calmness to capture a user's goals and state during agent execution. Because this grouping was constructed for this analysis and not validated as a measurement model, we treat all group-level comparisons as exploratory summaries of cross-category patterns rather than tests of a latent construct.

\begin{table*}
\centering
\scriptsize
\caption{The coding schema. The table lists six value groups constructed for this analysis, the 18 agent aspects in corpus-frequency order, and the nine user outcomes. Appendix~\ref{app:prompt-design} gives the operational definitions and boundary rules used in coding; Appendix~\ref{app:schema} adds a definition and an example excerpt for each of the 21 values.}
\label{tab:groups}
\begin{tabular}{@{}>{\raggedright\arraybackslash}p{2.5cm}>{\raggedright\arraybackslash}p{4.5cm}>{\raggedright\arraybackslash}p{2.6cm}>{\raggedright\arraybackslash}p{4.4cm}@{}}
\toprule
\textbf{Value group} & \multicolumn{2}{@{}>{\raggedright\arraybackslash}p{7.5cm}@{}}{\textbf{Definition}} & \textbf{Human values} {\scriptsize (Friedman et al.'s list~\cite{friedman2006vsd}; \textdagger\ study-specific, developed for this corpus following~\cite{ledantec2009values})} \\
\midrule
Dependable Operation & \multicolumn{2}{@{}>{\raggedright\arraybackslash}p{7.5cm}@{}}{Whether the agent's service holds and can be repaired} & dependability\textdagger, trust, repairability\textdagger \\
Autonomous Operation & \multicolumn{2}{@{}>{\raggedright\arraybackslash}p{7.5cm}@{}}{The user's own goals and state while the agent acts for them} & autonomy, human welfare, calmness \\
Affordable Operation & \multicolumn{2}{@{}>{\raggedright\arraybackslash}p{7.5cm}@{}}{The money and resources a run consumes} & affordability\textdagger, resource stewardship\textdagger, environmental sustainability \\
Bounded Reach & \multicolumn{2}{@{}>{\raggedright\arraybackslash}p{7.5cm}@{}}{What the agent may access and who controls that access} & security\textdagger, privacy, identity, property ownership, data sovereignty\textdagger, contextual integrity\textdagger \\
Reviewability & \multicolumn{2}{@{}>{\raggedright\arraybackslash}p{7.5cm}@{}}{Whether users can see, approve, and answer for what the agent does} & transparency\textdagger, meaningful human control\textdagger, accountability, informed consent \\
Equitable Access & \multicolumn{2}{@{}>{\raggedright\arraybackslash}p{7.5cm}@{}}{Who can use the agent at all} & universal usability, freedom from bias \\
\midrule
\textbf{Agent aspect} {\scriptsize (adapted from~\cite{wang2024survey,xi2025rise,sumers2023cognitive,parasuraman2000model,amershi2019guidelines,chen2025toward,su2025survey})} & \textbf{Definition} & \textbf{Agent aspect} & \textbf{Definition} \\
\midrule
Model core & The underlying LLM and how it is configured & Multi-agent orchestration & Coordination among agents and delegated workers \\
Resource accounting & What a run consumes and what it is billed for & Observability & Traces, logs, and visibility into what the agent did \\
System access & Getting the agent installed, authenticated, and ready to use & Human oversight & Approval gates and points where a user intervenes \\
Environment access & What the agent may reach once access is configured & Error handling & Detecting, reporting, and recovering from failed actions \\
Tool execution & Invoking tools, commands, and external operations & Other & A substantive agent-design concern outside the taxonomy \\
Action effects & User-visible changes the agent makes to files and data & Tool selection & Which tool or command the agent chooses \\
Task specification & The goal, scope, and stopping rules set for a run & Planning & Decomposing a task and forming or revising a plan \\
Memory & Context, retrieval, and what the agent keeps or forgets & Agent profile & The system prompt, persona, and framing instructions \\
Runtime performance & Latency, responsiveness, and stability under load & Reasoning & Reasoning the agent shows or the user reports \\
\midrule
\textbf{User outcome} {\scriptsize (adapted from~\cite{iso2018usability,hassenzahl2006ux,hart1988nasa,lee2004trust,parasuraman1997humans})} & \multicolumn{3}{@{}>{\raggedright\arraybackslash}p{11.6cm}@{}}{\textbf{Definition}} \\
\midrule
Task effectiveness & \multicolumn{3}{@{}>{\raggedright\arraybackslash}p{11.6cm}@{}}{Whether the agent completed the task and how correct or usable the result was} \\
Time efficiency & \multicolumn{3}{@{}>{\raggedright\arraybackslash}p{11.6cm}@{}}{Whether the agent saved the user time or cost them time} \\
Resource burden & \multicolumn{3}{@{}>{\raggedright\arraybackslash}p{11.6cm}@{}}{What a run cost the user in money and metered resources} \\
Supervision workload & \multicolumn{3}{@{}>{\raggedright\arraybackslash}p{11.6cm}@{}}{The effort of watching, checking, and managing what the agent did} \\
Affective response & \multicolumn{3}{@{}>{\raggedright\arraybackslash}p{11.6cm}@{}}{How the user felt about the experience} \\
Trust calibration & \multicolumn{3}{@{}>{\raggedright\arraybackslash}p{11.6cm}@{}}{How far the user relied on the agent, and whether that reliance was warranted} \\
Adoption behavior & \multicolumn{3}{@{}>{\raggedright\arraybackslash}p{11.6cm}@{}}{Whether the user continued with the agent, intended to, or moved away from it} \\
Risk exposure & \multicolumn{3}{@{}>{\raggedright\arraybackslash}p{11.6cm}@{}}{Harm to the user's data, systems, or compliance position that a post reported or clearly anticipated} \\
Recovery behavior & \multicolumn{3}{@{}>{\raggedright\arraybackslash}p{11.6cm}@{}}{What the user did to repair or work around what the agent did} \\
\bottomrule
\end{tabular}
\end{table*}

\subsection{Human Validation}
\label{sec:human_validation}
\textbf{Stage~1 validation.} We evaluated the relevance screen on a 50-post relevance pilot, distinct from the inductive sample in Section~\ref{sec:data_analysis}, against human consensus labels, achieving a precision of .900, recall of .783, and $F_1=.837$. Because these posts were not randomly sampled from the full corpus, these estimates do not establish full-corpus recall.

\textbf{Stage~2 validation.} We evaluated the Stage~2 codebook and initial LLM labels on the same 400-post sample used to develop the codebook (201 original posts, 199 comments), so these metrics do not estimate label error across the full corpus. Six coders across three pairs reviewed the source text and initial LLM labels, entering decisions independently before discussion (an LLM-assisted, rather than blinded, procedure). Appendix~\ref{app:human-validation} reports exact agreement, Cohen's $\kappa$~\cite{cohen1960coefficient,artstein-poesio-2008-survey}, accuracy, macro-precision, macro-recall, macro-$F_1$, and weighted-$F_1$. Overall, human--human agreement was consistently high (exact agreement = 82.0\%--98.8\%; $\kappa=.797$--$.979$), whereas LLM--human correspondence showed high accuracy (.900--.994) but more variable macro-$F_1$ (.604--.924) across coding targets.

\subsection{Statistical and Qualitative Analysis}
\label{sec:statistical_analysis}
\textbf{RQ1 analyses.} We summarized the 21 values and six value groups, then cross-tabulated the six groups with the 18 agent aspects. Pearson's $\chi^2$ summarized departure from independence and Cram\'er's $V$ its magnitude, while observed-to-expected (O/E) ratios described single cells, each comparing a value group's share at an agent aspect with that aspect's share of the whole corpus. We applied the same two measures to value group and value fulfillment. A value group's met rate reflected both its values and its agent aspects, because posts in different value groups raised different agent aspects, and because agent aspects differed in how often the values raised at them were met. We therefore used indirect standardization, comparing each value group's observed met rate with the rate expected if its posts had been met at the corpus-wide rate for the agent aspects they raised. Appendix~\ref{app:formal-measures} describes the equations, the bootstrap procedure behind the intervals we report, and what those intervals leave out. These standardized differences are descriptive.

\textbf{RQ2 analyses.} We cross-tabulated the six value groups with the nine user outcomes across the 44,767 outcome-coded posts, and applied the same two measures to value fulfillment and user outcome. Every expected count exceeded five in the tables we report (Appendix~\ref{app:formal-measures}). Each cell of the $6\times9$ table reports its post count and its met rate, the share of that cell's posts whose value was coded met. We displayed, but did not interpret, cells holding fewer than 20 posts, because in a cell that small a single post moves the met rate by more than five percentage points. 

\textbf{Thematic analysis of value-centered excerpts.} The cross-tabulations above report which values, agent aspects, and user outcomes co-occurred, but not how users described them, so we returned to the verbatim excerpt that grounded each coded post. We read these excerpts through an inductive thematic analysis~\cite{braun2006using}, where codes came from the excerpts rather than from a prior list, while reading stayed inside the dimensions the Stage~2 coding had already established. We treated each value group's two most frequent user outcomes as the unit of reading, because those twelve cells hold 69.9\% of the outcome-coded posts, and read a stratified random sample of 20 excerpts per cell, 240 excerpts in total. One researcher labeled what each excerpt claimed, grouped the labels into sub-themes, and grouped the sub-themes into themes, discussing the developing codebook with the research team throughout and revising labels and boundaries after each discussion. For example, the researcher labeled separately the posts reporting that a run ended without acting, that a fallback model answered without raising an error, and that finished-looking output had never been tested. Those labels were then collected into the sub-theme \textit{silent or partial failure} and paired with \textit{verification taken back by the user} to form the theme \textit{finishing a run stopped meaning the work was done}. The procedure yielded 63 initial codes, 24 sub-themes, and 12 themes, two per value group, which Sections~\ref{sec:rq2_autonomous}--\ref{sec:rq2_equitable} report as the bolded claims. Appendix~\ref{app:codebook} gives the codebook with a representative excerpt for each sub-theme. Because this step interpreted posts that the pipeline had already coded, it changed no count reported in this paper, and we report no interrater statistic for it.

\textbf{Unit of analysis.} The user post, not the user, was our unit of analysis. Because the dataset may contain multiple posts from a single user, our association statistics and bootstrap intervals do not adjust for within-user dependence and should be read as corpus-specific rather than population-level estimates. Every value and outcome we record also comes from what a user wrote about their own use rather than from an independent measure of the agent's behavior.

\subsection{Ethics}
\label{sec:ethics_availability}
\textbf{Research ethics.} Our university's institutional review board determined the study exempt because it analyzed publicly available posts without interacting with their authors. We reported aggregate patterns from posts and quoted only short excerpts reviewed for identifying details. To reduce reidentification risk, we removed usernames, links, timestamps, and community identifiers, and the publicly released research materials further excluded the verbatim source text.

\textbf{Use of AI tools.} We used a large language model as a coding instrument, as Section~\ref{sec:data_analysis} describes, and AI-based assistants for copy-editing and for checking references while preparing this manuscript. Following the ACM Policy on Authorship,\footnote{\url{https://www.acm.org/publications/policy-on-authorship}} no generative AI tool is listed as an author, every passage such a tool touched was reviewed by the authors, and the authors take full responsibility for the entire text.

\section{RQ1: What Human Values Surface in First-Person Posts of OpenClaw Use, and Where Do They Attach?}
\label{sec:rq1_results}
The RQ1 sample contained 73,093 posts. We first reported the distribution of human values (Section~\ref{sec:rq1_values}), then compared the agent-aspect distributions of the six value groups (Section~\ref{sec:rq1_aspects}). Finally, we examined value fulfillment by comparing each group's \emph{observed met rate}, the actual percentage of posts where the value was coded as successfully supported, with the rate expected given its distribution of agent aspects (Section~\ref{sec:rq1_fulfillment}).

\subsection{The Most Common Human Values Across Value Groups Were Autonomy, Dependability, Affordability, Resource Stewardship, and Universal Usability}
\label{sec:rq1_values}
\begin{figure*}[htbp]
  \centering
  \includegraphics[width=0.6\textwidth]{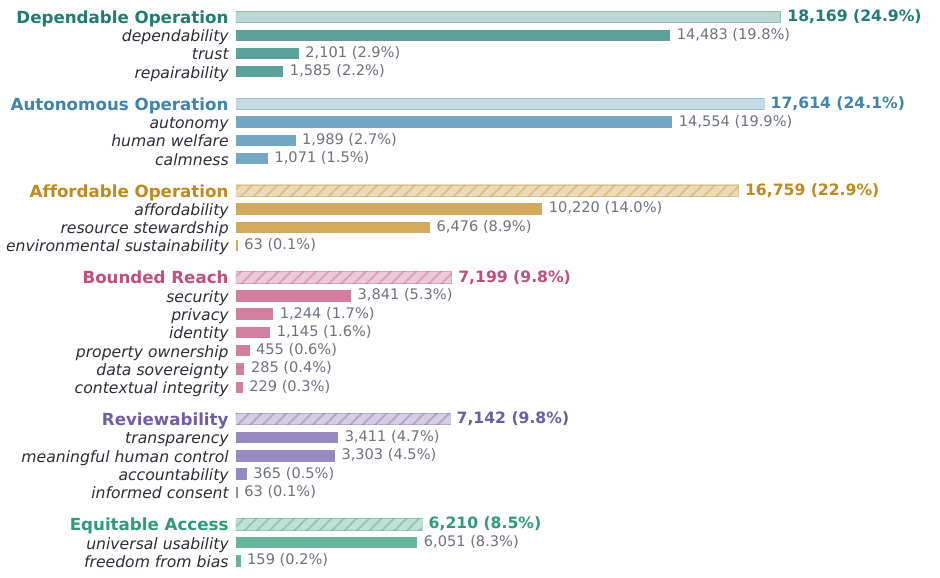}
  \caption{Distribution of values and their six value groups in the RQ1 sample ($N=73{,}093$). Solid bars show the number of posts of each value. The shaded group bars show the number of posts in each value group; the six group counts therefore sum to the RQ1 sample.}
  \Description{Bar chart showing 21 primary human values nested within six value groups. Autonomy and dependability are the two largest values, followed by affordability, resource stewardship, and universal usability. Dependable Operation, Autonomous Operation, and Affordable Operation are the three largest groups.}
  \label{fig:value_groups}
\end{figure*}

Five of the 21 values accounted for 51,784 of the 73,093 posts (70.8\%). Figure~\ref{fig:value_groups} shows all 21 values within their six groups. Specifically, autonomy was the most frequent value (19.9\%), followed by dependability (19.8\%), affordability (14.0\%), resource stewardship (8.9\%), and universal usability (8.3\%). Four of the remaining 16 values appeared in fewer than 250 posts each. We therefore reported the full value distribution but used the six value groups for cross-category comparisons, which would otherwise include value-level cells with few posts.

The three largest value groups, Dependable Operation, Autonomous Operation, and Affordable Operation, contained 52,542 posts (71.9\%); Bounded Reach, Reviewability, and Equitable Access contained the remaining 20,551 posts. User posts thus focused more on whether the OpenClaw agent worked, what it could do on its own, and what it cost, and less often on what it could reach, whether users could review it, or who could use it. By provenance, the 12 VSD-derived values accounted for 29,260 posts (40.0\%) and the nine study-specific values for 43,833 (60.0\%).

\subsection{Four Value Groups Were Each Associated with One or Two Agent Aspects, Whereas the Two Largest Shared One Aspect, the Model Core}
\label{sec:rq1_aspects}

\begin{figure*}[htbp]
  \centering
  \includegraphics[width=0.6\textwidth]{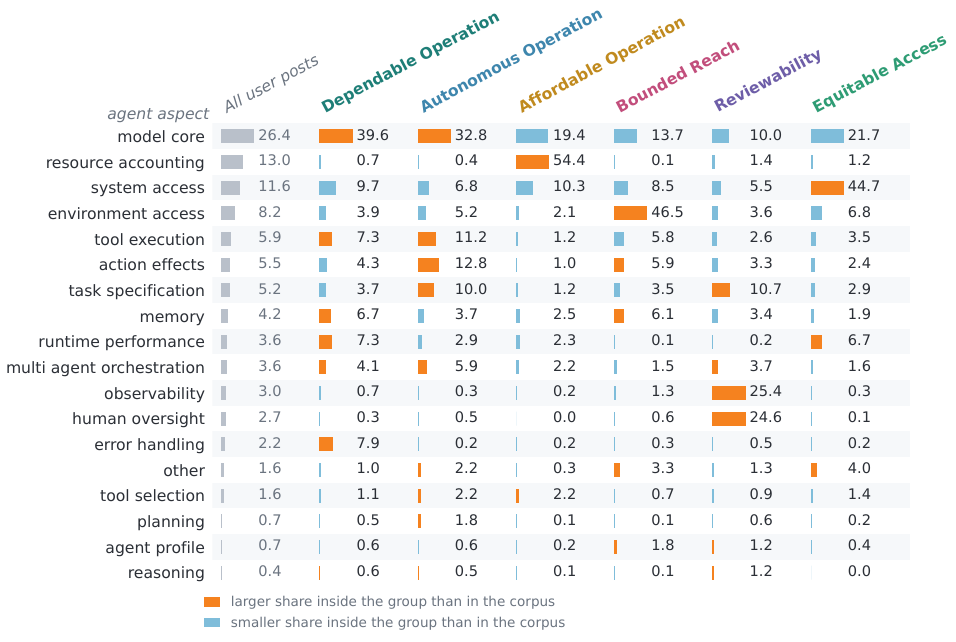}
  \caption{Distribution of agent aspects across the six value groups in the RQ1 sample ($N=73{,}093$). Each cell shows the percentage of posts in a given value group coded for that agent aspect, and each group column sums to 100\%. The first column gives each aspect's share of the whole corpus; orange marks a larger share inside the group than in the corpus (O/E ratio above 1) and blue a smaller share, and Appendix~\ref{app:formal-measures} defines the ratio.}
  \Description{Matrix of 18 agent aspects by six value groups. Affordable Operation is concentrated on resource accounting, Bounded Reach on environment access, Equitable Access on system access, and Reviewability on observability and human oversight. Dependable Operation and Autonomous Operation both most often concern the model core, but differ in several less frequent aspects.}
  \label{fig:agent_aspect_composition}
\end{figure*}

Value group and agent aspect were associated at the post level, $\chi^2(85, N=73{,}093)=94{,}536.8$, with Cram\'er's $V=.509$. Figure~\ref{fig:agent_aspect_composition} reports each value group's agent-aspect distribution beside the corresponding corpus-wide share.

Three value groups each concentrated on a single agent aspect. Resource accounting comprised 54.4\% of Affordable Operation posts against 13.0\% corpus-wide (O/E 4.2), environment access 46.5\% of Bounded Reach posts against 8.2\% (O/E 5.7), and system access 44.7\% of Equitable Access posts against 11.6\% (O/E 3.9). Read the other way, those three value groups differ sharply. Affordable Operation held 96.1\% of all resource-accounting posts, Bounded Reach 55.8\% of environment-access posts, and Equitable Access only 32.8\% of system-access posts. Cost was therefore discussed almost exclusively within Affordable Operation, whereas user setup and authentication arose across groups rather than belonging to Equitable Access.

Reviewability rested on two agent aspects rather than one, observability at 25.4\% of its posts (O/E 8.6) and human oversight at 24.6\% (O/E 9.2). Reviewability held 83.9\% of all observability posts and 89.9\% of all human-oversight posts. Seeing what the OpenClaw agent had done and being asked to approve it therefore surfaced together, and almost only where users raised Reviewability.

The two largest value groups shared an agent aspect. Both discussed the model core most often, at 39.6\% of Dependable Operation posts and 32.8\% of Autonomous Operation posts. The model core was also the most common agent aspect corpus-wide at 26.4\%, so neither value group was more than modestly overrepresented there (O/E 1.5 and 1.2). The two value groups separated below it. Dependable Operation reached O/E ratios of 3.7 for error handling and 2.0 for runtime performance, whereas Autonomous Operation reached 2.3 for action effects and 1.9 for both tool execution and task specification. The agent aspect a value group discussed most was therefore not always the agent aspect that distinguished it. These ratios describe cells within the overall association, not separate cellwise tests.

\subsection{Value Fulfillment Rate Varied Across Value Groups, but Only Autonomous and Affordable Operation Exceeded Expected Rates}
\label{sec:rq1_fulfillment}
\begin{figure*}[htbp]
  \centering
  \includegraphics[width=0.6\textwidth]{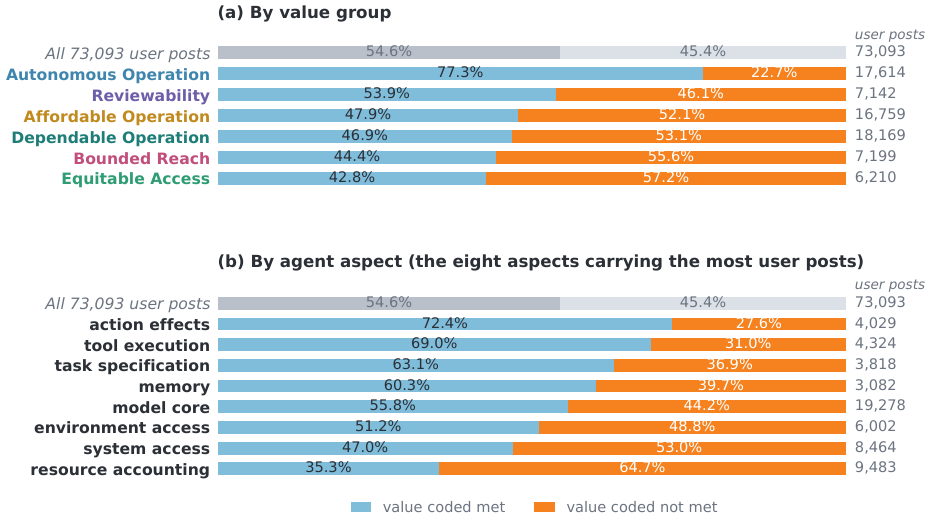}
  \caption{Value fulfillment in the RQ1 sample ($N=73{,}093$). Panel (a) reports met and not-met shares for the six value groups. Panel (b) reports the same shares for the eight most frequent agent aspects.}
  \Description{Two stacked bar charts. Autonomous Operation has the highest group-level met rate at 77.3 percent, and Equitable Access has the lowest at 42.8 percent. Among the eight most frequent agent aspects, action effects have the highest met rate and resource accounting the lowest.}
  \label{fig:value_fulfillment}
\end{figure*}

Across the RQ1 sample, the value at stake was coded met in 39,878 of 73,093 posts (54.6\%). Met rates ranged from 77.3\% for Autonomous Operation to 42.8\% for Equitable Access, with 53.9\% for Reviewability, 47.9\% for Affordable Operation, 46.9\% for Dependable Operation, and 44.4\% for Bounded Reach (Figure~\ref{fig:value_fulfillment}(a)). Also, value group and fulfillment were associated at the post level, $\chi^2(5, N=73{,}093)=5{,}056.4$, with Cram\'er's $V=.263$.

Dependable Operation and Affordable Operation together accounted for 18,374 of the 33,215 not-met posts (55.3\%). A high group met rate also did not mean uniform fulfillment inside a value group, since autonomy, the most frequent value, still had 2,208 of its 14,554 posts (15.2\%) coded not met, 6.6\% of all not-met posts. Across the 18 agent aspects, met rates ranged from 81.3\% for planning (401 of 493 posts) to 35.3\% for resource accounting (3,347 of 9,483). Figure~\ref{fig:value_fulfillment}(b) shows the eight agent aspects with the most posts. Value fulfillment thus varied more widely across agent aspects than across value groups, so part of the spread between value groups could reflect which agent aspects their posts raised (Section~\ref{sec:rq1_aspects}) rather than the values themselves.

\begin{figure*}[htbp]
  \centering
  \includegraphics[width=0.6\textwidth]{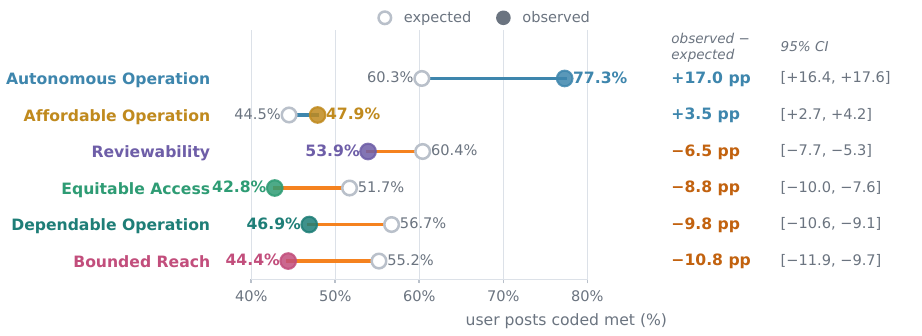}
  \caption{Observed value fulfillment and fulfillment expected from the distribution of agent aspects in each group in the RQ1 sample ($N=73{,}093$). Differences are observed minus expected percentage points; intervals are conditional 95\% percentile bootstrap intervals from 2,000 post-level resamples within each value group.}
  \Description{Dot-and-interval plot showing observed-minus-expected fulfillment for six value groups. Autonomous Operation is 17.0 pp above expectation, and Affordable Operation is 3.5 pp above. Reviewability, Equitable Access, Dependable Operation, and Bounded Reach are below expectations.}
  \label{fig:actual_vs_expected_fulfillment}
\end{figure*}

We therefore applied the indirect standardization described in Section~\ref{sec:statistical_analysis} (Figure~\ref{fig:actual_vs_expected_fulfillment}). Observed minus expected was positive for two value groups, $+17.0$ pp for Autonomous Operation (conditional 95\% percentile bootstrap interval [$+16.4$, $+17.6$]) and $+3.5$ for Affordable Operation ([$+2.7$, $+4.2$]). It was negative for the other four, $-6.5$ for Reviewability ([$-7.7$, $-5.3$]), $-8.8$ for Equitable Access ([$-10.0$, $-7.6$]), $-9.8$ for Dependable Operation ([$-10.6$, $-9.1$]), and $-10.8$ for Bounded Reach ([$-11.9$, $-9.7$]). These intervals resample posts within each value group with corpus-wide agent-aspect-specific met rates held fixed, and describe the coded corpus rather than causal or user-level effects.

A positive difference means a value group met its values more often than the agent aspects it discussed would predict. Affordable Operation ranked third on observed met rate yet exceeded its expectation, because 54.4\% of its posts concerned resource accounting, the least-met agent aspect in the corpus at 35.3\%. Affordable Operation's shortfall therefore lay in the agent aspect its posts raised rather than in the values at stake. Reviewability moved the other way, its second-highest observed met rate falling short of an expectation set by two agent aspects met more often than the corpus-wide 54.6\%, observability at 56.4\% and human oversight at 67.7\%. Autonomous Operation's advantage survived the adjustment, at 77.3\% observed against 60.3\% expected, so it did not follow from the agent aspects its posts raised.

\section{RQ2: What Outcomes Do Users Attribute to Agent Use Across Value Groups?}
\label{sec:rq2_results}
The RQ2 subset contained 44,767 of the 73,093 posts (61.2\%) that carried an attributable user outcome. Among the nine user-outcome categories (Table~\ref{tab:groups}), task effectiveness was the most frequent (18,646 posts, 41.7\%), followed by resource burden (6,955, 15.5\%) and time efficiency (5,332, 11.9\%). Value group and user outcome were associated at the post level, $\chi^2(40, N=44{,}767)=41{,}092.5$, with Cram\'er's $V=.428$. Figure~\ref{fig:value_outcome_fulfillment_heatmap} reports that $6\times9$ cross-tabulation. To make the six value-group summaries comparable, each subsection below reports the two most frequent user outcomes and describes how users characterized them. A closing subsection then reads the same posts across value groups, describing how often values were met in each user outcome and user outcomes when a value was met and not (Section~\ref{sec:rq2_fulfillment}).

\subsection{Autonomous Operation: Reach Users Widened Rather Than Took Back}
\label{sec:rq2_autonomous}
Task effectiveness was the most frequent user outcome for Autonomous Operation (5,285 of the 11,754 outcome-coded posts in this group, 45.0\%), followed by time efficiency (2,823 posts, 24.0\%). 

\textbf{Users kept widening what the agent could touch.} One user with no coding knowledge asked how to make OpenClaw do what they had seen others do, describing ``\textit{letting it open programs, searching the web, actually doing work autonomously.}'' Another described what they were adding next: ``\textit{I'm working on getting mine a phone number, and access to a debit card with access to capital.}'' In both cases, users widened what OpenClaw could reach on their devices and accounts rather than restricting it.

\textbf{Users got time back, not better output.} When users said what they had gained from using OpenClaw, they often described being freed from doing the work themselves rather than getting a better result. One user wrote, ``\textit{Giving OpenClaw a pair of hands has freed up my hands!}'' Time efficiency here recorded the hours the user no longer had to spend, not a task finished faster. Another user set the promise of returned time against the wider situation it arrived in: \textit{``OpenClaw was supposed to give us more time, but it feels like time is decreasing by the day. What's the point of everything if no one uses it? Building is easy now, but who will purchase, given that building is easy? How will supply and demand match?''} This user questioned what their own effort was worth once task execution became easy.

\subsection{Dependable Operation: Delivery That Did Not Hold Across Runs}
\label{sec:rq2_dependable}
Dependable Operation's posts led with task effectiveness (8,993 of its 14,256 outcome-coded posts, 63.1\%), followed by adoption behavior (1,249 posts, 8.8\%). The remaining user outcomes were sparse. Dependable Operation posts stood apart because users described the same OpenClaw setup as delivering at one time and failing to keep delivering at another, which left them asking whether the agent would carry out a task again.

\textbf{Finishing a run stopped meaning the work was done.} Users expected an instruction to become an OpenClaw action, and posts about failure described the agent stopping short of that step rather than producing a wrong result. One user described an agent that would not act on a plan it had itself proposed: ``\textit{No matter how many times I approve, it never actually transitions from `planning' to `doing.'}\,'' Failure was also not always visible at the moment it occurred, since free models, as one user put it, ``\textit{don't always fail loudly}'' and would ``\textit{quietly ship you a stub and move on.}'' Because a finished run and an empty run looked alike, task completion stopped serving as evidence that the work had been done, and users took verification back. One user, describing repeated breakage, wrote: \textit{``I kept breaking them in production almost daily because I treated OpenClaw changes like scratch work instead of production deployments.''} This user treated OpenClaw configuration as a deployment problem rather than a settings change, and keeping that configuration working was itself infrastructure work, the kind of task OpenClaw was meant to remove. Where a setup did hold, the posts were correspondingly plain, as one user who had helped others debug their OpenClaw configurations observed: ``\textit{the setups that survive are the ones where the person can explain what their agent does in one sentence.}''

\textbf{Users kept the agent but asked less of it.} Repeated failure rarely ended agent use. One user planned to ``\textit{uninstall Qwen 9.5 9b today and give Gemma a shot,}'' and to ``\textit{switch to Ollama cloud}'' if the replacement also fell short. Another judged that ``\textit{the openclaw product became undesirable at least compared to other platform options}'' and chose to downgrade and wait until it improved, rather than to stop. After a failure, these users changed the model, the host, or the task's ambition, not the decision to delegate. Adoption behavior in this value group therefore recorded lowered expectations; the OpenClaw agent stayed, but users stopped counting on it to behave the same way twice.

\subsection{Affordable Operation: Cost That Accrued Where Users Could Not See It}
\label{sec:rq2_affordable}
Resource burden was the most frequent user outcome for Affordable Operation (6,776 of the 10,549 outcome-coded posts, 64.2\%), followed by task effectiveness (1,359 posts, 12.9\%). These two user outcomes did not offset each other. A run that completed a task could still generate a bill, and the bill arrived without explaining which agent actions had cost money, so posts in this value group registered OpenClaw's usefulness and its resource consumption as separate primary outcomes.

\textbf{The bill came from work users never asked for.} What these posts objected to was rarely that a price was too high, but that consumption happened where the user was not looking. One user found they had been ``\textit{paying Sonnet rates for 48 heartbeats a day to silently check if anything was scheduled,}'' a charge produced by the agent's readiness rather than by any task they had asked for. Another traced the same problem to what each message loaded, proposing that ``\textit{if the user says `hello', don't inject the full system prompt and skill definitions.}'' A third user added: \textit{``OpenClaw is not being proactive in monitoring token usage and preventing input tokens from reaching higher than 100K per message, even when I'm only writing `test' and with things in place to alert at 40K input tokens to compact and reduce back down to sub-20K.''} This user had already built the accounting that the agent lacked, including an alert threshold and scheduled sessions to compact context, and still could not keep consumption down. What they described was a spending decision made inside a run rather than at its boundary. Cost in this value group, therefore, behaved less like a price than like a running cost, visible after the fact but not available to plan against.

\textbf{Users moved the spending rather than stopping it.} Posts that described cost as settled rarely described a cheaper agent. They described moving the spending somewhere the user could predict it. One user compared plans directly and got ``\textit{more usage out of GitHub Copilot Pro+ for \$40/month using Claude models than I do from 5x Claude Team accounts for \$100.}'' Others changed the agent rather than the plan, adopting a minimal agent that consumed fewer tokens under default prompts, which one user reported had become ``\textit{the only coding agent I use now.}'' Where the accounting did settle, users measured the agent against the tools it displaced. One described applications an agent had built that ``\textit{I previously would have had to pay anywhere from 20 to 200 dollars a month for.}'' Affordability in this value group was accordingly not a property of the agent but an arrangement users assembled around it, and that arrangement lasted only while they kept adjusting it.

\subsection{Bounded Reach: Risk and Usefulness Both Settled at OpenClaw Setup}
\label{sec:rq2_bounded}

Risk exposure was the most frequent user outcome for Bounded Reach (719 of the 2,425 outcome-coded posts in this group, 29.6\%), followed by task effectiveness (585 posts, 24.1\%). Users described the two in different terms. \textbf{The risk users described was access, not damage.} Posts with this user outcome described the access the user had granted rather than harm the user had suffered. That access was a property of an OpenClaw configuration, meaning which files, credentials, and networks the agent could reach, and users fixed it when they installed the agent rather than during a task. One user weighing a business deployment was ``\textit{afraid to connect it to my company's data and tools.}'' Another listed ``\textit{whether it can touch anything outside the workspace indirectly.}'' Users understood the permission model as an instruction set, and as one put it, the configuration was ``\textit{a set of instructions your agent follows with full permissions.}'' Another user described a default that shipped open: \textit{``OpenClaw has had 14 CVEs [Common Vulnerabilities and Exposures] since launch, 8 of them critical. The default config binds the gateway to 0.0.0.0, which means anyone who finds your IP has full access to your OpenClaw admin. You need to bind to localhost, set up sandbox mode, configure tool deny lists, and add systemd isolation.''} This user separated the discovered defects from a setting that shipped open and listed four things the operator had to add. The gateway was reachable not because a run went wrong but because that was how the software shipped, so the risk lay in how OpenClaw was set up rather than in what a run produced.

\textbf{Useful runs happened inside limits set beforehand.} The posts described not an agent given more room but a setup whose reach the user had already set in advance. One user reported a placement, having ``\textit{launched this [OpenClaw] on LightSail to avoid it having access to my files.}'' Another stated a data boundary, ``\textit{I'll use my local inference models for internal stuff only.}'' Users described tasks that went well after such configuration. One noticed that ``\textit{it keeps memory separated by project}'' and found this ``\textit{helpful when organizing a lot of scattered notes and materials.}''

\subsection{Reviewability: Checkpoints That Worked at a Run's Edges but Not Inside It}
\label{sec:rq2_reviewability}
Task effectiveness was the most frequent user outcome for Reviewability (982 of the 2,590 outcome-coded posts in this group, 37.9\%), followed by affective response (492 posts, 19.0\%). 
\textbf{Checking before or after a run cost nothing.} Where users described OpenClaw's work as useful, they had set the point of control before the run or reviewed the outcome after it. One user who had delegated guest-post outreach reported that the agent ``\textit{didn't just say yes to everything}'' and ``\textit{pushed back}'' on a topic outside their expertise. That refusal came from instructions the user had given in advance, not from a question the user answered during the run. Another user kept one standing gate on the outbound run, ``\textit{nothing sends without a human},'' and reduced their involvement to that single decision. 
In these posts, what users counted as success was the outcome itself, such as the outreach that closed. Reviewing the agent asked little of these users because none of it happened while the run was executing.

\textbf{The cost was in answering, not in looking.} What users objected to was not that the agent asked for approval, but that it asked about a step they had already approved or asked somewhere they could not answer. One user asked, \textit{``How can I make it so that my Mac Mini doesn't continue to request permissions to do things I am asking it to do? It continues to interrupt the workflow, and I feel like I haven't set things up correctly.''} This user had authorized the steps, and the agent asked again. They read OpenClaw's repeated request not as oversight but as evidence that their own setup was wrong. Another user, running OpenClaw on a headless node, reported that ``\textit{the approval dialogs never actually appear,}'' and kept a browser window open to authorize each step by hand. The gate existed, but it was positioned out of reach. A third user, after switching to a different frontend, could ``\textit{actually see the tool calls and reasoning steps in a clean workspace}'' and called the view ``\textit{a huge sanity saver.}'' Such a view let the user monitor the agent without interruption. What separated a workable checkpoint from an interruption in this value group was not whether users could watch the agent, but whether watching required an answer during the run.

\subsection{Equitable Access: Useful Work Reported From the Far Side of Setup}
\label{sec:rq2_equitable}
Task effectiveness (1,442 of the 3,193 outcome-coded posts, 45.2\%) and adoption behavior (585 posts, 18.3\%) were the two most frequent user outcomes for Equitable Access. These posts came from two kinds of users, those who had already reached a working OpenClaw setup and those still describing what it cost to reach one. 

\textbf{Posts of useful work came from users already past setup.} One user had cleared setup and moved into a threaded chat client, reporting that ``\textit{I can chat with my agent in different threads at the same time, and everything stays nice and neat.}'' When users explained how they reached a working setup, they credited help that someone else had prepared. One user switching providers followed a documented onboarding command and reported that ``\textit{the change was pretty easy.}'' Another user, running OpenClaw on a hosting provider, credited the provider rather than the agent, writing that ``\textit{Contabo has made it really easy to get something like OpenClaw up, I've started messing with it.}''

\textbf{Users weighed the install before they weighed the agent.} Posts about adoption behavior often recorded a decision made before or during setup rather than after extended use. One user packaging a standalone build reasoned that ``\textit{not everyone is comfortable using command line or Docker on Linux,}'' and another asked, before installing anything, ``\textit{Openclaw for non-techie.}'' Completing the install did not settle the question, as one user who had finished it called the process ``\textit{still annoying as hell to setup.}'' A third user added: \textit{``Accessibility is what made OpenClaw popular. Not the accessibility of OpenClaw itself, but what it allows non/moderately technical people to achieve, and what real-world problems they could solve with it.''} This user separated reaching the agent from what the agent then made achievable, and credited OpenClaw only with the second. 

\subsection{Values Were Met More Often Where Users Described Delivery Than Where They Described Cost}
\label{sec:rq2_fulfillment}
\begin{figure*}[htbp]
  \centering
  \includegraphics[width=0.65\textwidth]{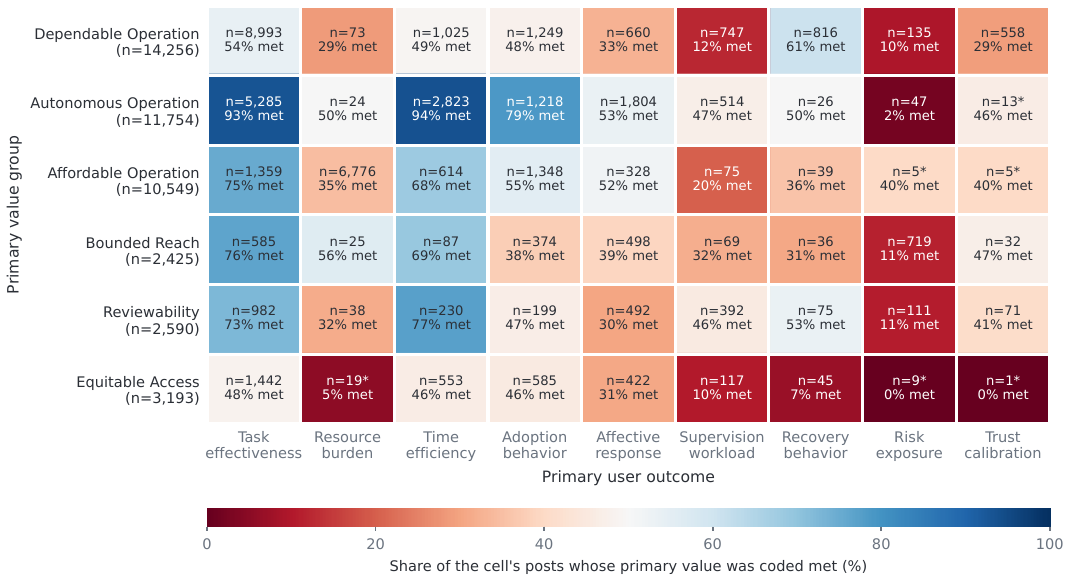}
  \caption{Value group by user outcome among the 44,767 outcome-coded posts. Each cell describes the post count and the met rate, the share of its posts whose human value was coded as met. Blue is majority met, red is majority not met, and cells at 50\% are unshaded. Asterisks mark cells with fewer than 20 posts, shown but not interpreted.}
  \Description{Heatmap with six value groups as rows and nine user outcomes as columns. Task effectiveness is majority met in five of the six value groups, most strongly in Autonomous Operation. Supervision workload is majority not met in every group, and risk exposure is majority not met in every cell with at least 20 posts. Affordable Operation contains most resource-burden posts, and Bounded Reach contains the largest risk-exposure cell.}
  \label{fig:value_outcome_fulfillment_heatmap}
\end{figure*}

Among the 44,767 outcome-coded posts, the value was met in 25,172 (56.2\%) and not met in 19,595 (43.8\%). Figure~\ref{fig:value_outcome_fulfillment_heatmap} breaks that split down by value group and user outcome, giving the post count and the met rate in each of its 54 cells. These cell rates describe the outcome-coded subset and are therefore not directly comparable with the group met rates in Section~\ref{sec:rq1_fulfillment}, which cover all 73,093 coded posts. The share of a value group's posts carrying an attributable user outcome also varied, from 33.7\% in Bounded Reach to 78.5\% in Dependable Operation, so the value groups are not equally represented in this subset.

Whether a value held depended more on the user outcome a post reported than on its value group. The value was met in 67.7\% of the 18,646 task-effectiveness posts and in 76.1\% of the 5,332 time-efficiency posts. The value was met in only 34.7\% of the 6,955 resource-burden posts, 29.4\% of the 1,914 supervision-workload posts, and 10.7\% of the 1,026 risk-exposure posts. Task effectiveness was majority-met in five of the six value groups, from 93.0\% in Autonomous Operation down to 53.6\% in Dependable Operation; only in Equitable Access, at 48.4\%, did its posts fall close to even without clearing half. Supervision workload ran the other way in all six, from 46.9\% met in Autonomous Operation down to 10.3\% in Equitable Access. Risk exposure ran the same way but fell below the 20-post display threshold in two value groups, so supervision workload was the only majority-not-met user outcome whose cells cleared that threshold in every value group.

Three value groups paired one majority-met and one majority-not-met leading user outcome. Bounded Reach showed the widest gap, with the value met in 76.4\% of its task-effectiveness and in 11.4\% of its risk-exposure posts. Reviewability followed, with task effectiveness at 72.6\% and affective response at 30.1\%, then Affordable Operation, with task effectiveness at 75.1\% and resource burden at 34.7\%. Each post carried one user outcome, so these opposite directions came from different sets of posts within the same value group rather than from single users. In the other three value groups, the two leading user outcomes sat close together, high for Autonomous Operation (93.0\% and 93.6\%) and nearly even for Dependable Operation (53.6\% and 48.0\%) and Equitable Access (48.4\% and 45.6\%). Particular user outcomes also concentrated in particular value groups, with Affordable Operation holding 6,776 of the 6,955 resource-burden posts (97.4\%) and Bounded Reach holding 719 of the 1,026 risk-exposure posts (70.1\%).

Setting the value groups aside and cross-tabulating value fulfillment against user outcome shows which user outcomes users reported on each side of fulfillment, $\chi^2(8, N=44{,}767)=5{,}033.9$, with Cram\'er's $V=.335$ (Table~\ref{tab:fulfillment_outcome}). Among met posts, task effectiveness and time efficiency together accounted for 66.2\%, and both name what a run gave back rather than what it cost. Among not-met posts those two user outcomes accounted for 37.3\%. Four user outcomes ran the other way, with resource burden, affective response, supervision workload, and risk exposure accounting for 46.9\% of not-met posts against 19.5\% of met posts, and each of those four recorded what running the agent cost. Adoption behavior did not follow that contrast, standing at 2,802 met and 2,171 not-met posts (11.1\% of each), though the six value groups differed on it. These were post-level associations among one coded label per field and did not establish causal effects, estimates for unique users, or co-occurring user outcomes within a post.

\begin{table*}[htbp]
\centering
\scriptsize
\setlength{\tabcolsep}{8pt}
\caption{User outcomes among the 44,767 outcome-coded posts, split by whether the post's value was coded met. Rows sum to 100\% before rounding, and the final row gives the met-minus-not-met difference in percentage points. 
}
\label{tab:fulfillment_outcome}
\begin{tabular}{@{}l*{9}{r}r@{}}
\toprule
 & \textbf{\shortstack{Task\\eff.}} & \textbf{\shortstack{Resource\\burden}} & \textbf{\shortstack{Time\\eff.}} & \textbf{\shortstack{Adoption\\behavior}} & \textbf{\shortstack{Affective\\response}} & \textbf{\shortstack{Supervision\\workload}} & \textbf{\shortstack{Recovery\\behavior}} & \textbf{\shortstack{Risk\\exposure}} & \textbf{\shortstack{Trust\\calibration}} & \textbf{$n$} \\
\midrule
Value met & 50.1 & 9.6 & 16.1 & 11.1 & 7.2 & 2.2 & 2.3 & 0.4 & 0.9 & 25,172 \\
Value not met & 30.8 & 23.2 & 6.5 & 11.1 & 12.2 & 6.9 & 2.3 & 4.7 & 2.4 & 19,595 \\
\textit{Difference (pp)} & $+19.3$ & $-13.6$ & $+9.6$ & $0.0$ & $-5.0$ & $-4.7$ & $0.0$ & $-4.3$ & $-1.5$ & \\
\bottomrule
\end{tabular}
\end{table*}

\section{Discussion}
While prior work evaluates agents by task completion~\cite{liu2024agentbench,jimenez2024swebench,yao2025taubench} or scores value-relevant judgments~\cite{wang2024fakealignment,kirk2024prism}, our study is among the first to trace, using OpenClaw as an exemplar agent, which human values users invoke in everyday agent use and which agent aspects those values attach to. Below, we describe this relationship as \emph{value-sensitive delegation} and ground it in VSD's interactional position (Section~\ref{sec:disc_delegation}), examine the gains and costs users attributed to running OpenClaw (Section~\ref{sec:disc_gains_costs}), and translate both into design implications (Section~\ref{sec:disc_design}).

\definecolor{f7mut}{HTML}{4A4A46}
\definecolor{f7line}{HTML}{D8D7D2}
\definecolor{f7acc}{HTML}{B4560C}
\definecolor{f7u}{HTML}{2F6F9E}   \definecolor{f7ut}{HTML}{EEF4F9}
\definecolor{f7g}{HTML}{6A5AA6}   \definecolor{f7gt}{HTML}{F2F0F8}
\definecolor{f7p}{HTML}{1F7A6B}   \definecolor{f7pt}{HTML}{E9F4F1}
 
\newcommand{\ficon}[4]{%
  \begin{scope}[shift={(#1,#2)},scale=0.0145,shift={(-10,-10)},
                draw=#3,fill=#3,line width=0.5pt,line cap=round,line join=round]#4\end{scope}}
\newcommand{\ipers}{\draw (10,14) circle (3.4); \draw (3.6,2.6) arc (180:0:6.4);}
\newcommand{\ibot}{\draw[rounded corners=2] (2.5,3) rectangle (17.5,14);
  \fill (7,9.5) circle (1.3); \fill (13,9.5) circle (1.3);
  \draw (7.5,6)--(12.5,6); \draw (10,14)--(10,17); \fill (10,18.2) circle (1.2);}
\newcommand{\ibolt}{\draw (11.5,19)--(4.5,9.5)--(9.5,9.5)--(8.5,1)--(15.5,11)--(10.5,11)--cycle;}
\newcommand{\iloop}{\draw (10,17) arc (90:-215:7); \fill (2.2,15.6)--(7.2,16.4)--(4.4,11.6)--cycle;}
\newcommand{\icoin}{\draw (10,10) circle (8); \draw (10,3.6)--(10,16.4);
  \draw (5.8,12.8)--(14.2,12.8); \draw (5.8,7.2)--(14.2,7.2);}
\newcommand{\ilock}{\draw[rounded corners=1.6] (3,2) rectangle (17,11);
  \draw (6.2,11)--(6.2,14) arc (180:0:3.8) --(13.8,11);}
\newcommand{\ieye}{\draw (1.8,10) .. controls (6,16.8) and (14,16.8) .. (18.2,10)
  .. controls (14,3.2) and (6,3.2) .. (1.8,10) -- cycle; \draw (10,10) circle (3);}
\newcommand{\idoor}{\draw (5,3) rectangle (15,18); \fill (12.7,10) circle (1.1);
  \draw (1.5,3)--(18.5,3);}
\newcommand{\irec}{\draw (5,18)--(15,18)--(15,3)--(13,4.4)--(11,3)--(9,4.4)--(7,3)--(5,4.4)--cycle;
  \draw (7.6,14)--(12.4,14); \draw (7.6,10.6)--(12.4,10.6);}
\newcommand{\idoc}{\draw (5,18)--(12,18)--(16,14)--(16,3)--(5,3)--cycle;
  \draw (12,18)--(12,14)--(16,14); \draw (7,10)--(13,10); \draw (7,7)--(13,7);}
\newcommand{\iglass}{\draw (8.6,12) circle (5.4); \draw (12.6,8.1)--(17.6,3.1);}
\newcommand{\icheck}{\draw (10,10) circle (8); \draw (6,10.4)--(9,7.4)--(14.4,13.4);}
\newcommand{\igauge}{\draw (2,6) arc (180:0:8); \draw (10,6)--(14.6,12.6); \fill (10,6) circle (1.3);}
 
\begin{figure*}[t]
\centering
\resizebox{\textwidth}{!}{%
\begin{tikzpicture}[
  x=1cm,y=1cm,
  every node/.style={inner sep=0pt,outer sep=0pt},
  f7hd/.style={font=\normalsize\bfseries,anchor=west},
  f7sb/.style={font=\footnotesize,text=f7mut,anchor=west},
  f7cl/.style={font=\small,text=f7mut,anchor=west},
  f7gp/.style={font=\normalsize,anchor=west},
  f7cg/.style={font=\normalsize,anchor=west},
  f7cx/.style={font=\normalsize,text=f7acc,anchor=west},
  f7as/.style={font=\small,text=f7mut,anchor=west},
  f7ax/.style={font=\small,text=f7acc,anchor=west},
  f7sc/.style={font=\footnotesize,text=f7mut,anchor=east},
  f7lg/.style={font=\footnotesize,text=f7mut,anchor=west},
  f7bn/.style={font=\footnotesize\bfseries,text=white,anchor=center},
  f7al/.style={font=\footnotesize,text=f7mut,anchor=west},
  f7bl/.style={font=\normalsize\bfseries,text=f7acc,anchor=east},
  f7rl/.style={draw=f7line,line width=0.5pt},
  f7rl2/.style={draw=f7line,line width=0.35pt},
  f7arw/.style={draw=f7mut,line width=0.7pt,-{Stealth[length=5pt,width=3.6pt]}},
  f7dsh/.style={draw=f7mut,line width=0.55pt,dash pattern=on 2.4pt off 1.8pt,-{Stealth[length=5pt,width=3.6pt]}},
  f7bak/.style={draw=f7acc,line width=1.2pt,-{Stealth[length=6.5pt,width=5pt]}},
]
\fill[white,draw=f7u,line width=0.7pt,rounded corners=6pt] (0.4,-4.94) rectangle (19.0,0);
\fill[f7u,draw=white,line width=1.1pt] (0.8600000000000001,-0.45) circle (0.21);
\node[f7bn] at (0.8600000000000001,-0.45) {1};
\ficon{1.3}{-0.45}{f7u}{\ipers}
\node[f7hd,text=f7u] at (1.58,-0.45) {Before an OpenClaw \includegraphics[width=0.4cm]{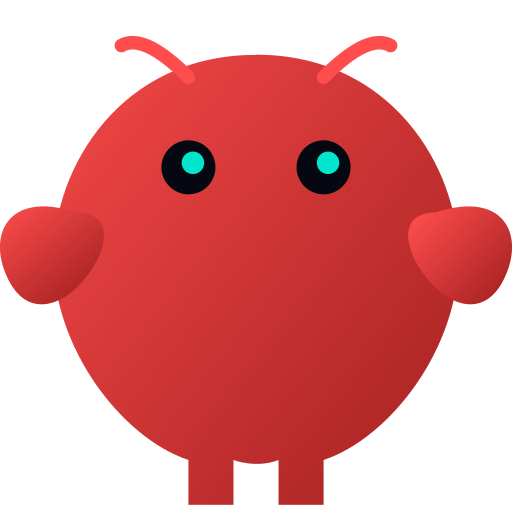} Agent Run};
\node[f7sb] at (0.78,-0.93) {the conditions a user set in advance};
\node[f7cl] at (1.3,-1.40) {Value Group};
\node[f7cl] at (5.1,-1.40) {Most Discussed Agent Aspect};
\node[f7cl] at (9.4,-1.40) {The Condition};
\node[f7cl] at (13.1,-1.40) {When Users Could Change It};
\draw[f7rl] (0.78,-1.58) -- (18.82,-1.58);
\ficon{0.9500000000000001}{-1.940}{f7u}{\ibolt}
\node[f7gp] at (1.3,-1.940) {Autonomous Operation};
\node[f7ax] at (5.1,-1.940) {model core};
\node[f7cx] at (9.4,-1.940) {the reach};
\node[f7cg] at (13.1,-1.940) {before a later run};
\node[f7sc] at (18.82,-1.940) {\S\ref{sec:rq2_autonomous}};
\draw[f7rl2] (0.78,-2.170) -- (18.82,-2.170);
\ficon{0.9500000000000001}{-2.400}{f7u}{\iloop}
\node[f7gp] at (1.3,-2.400) {Dependable Operation};
\node[f7as] at (5.1,-2.400) {model core};
\node[f7cg] at (9.4,-2.400) {the model and host};
\node[f7cg] at (13.1,-2.400) {after a failure, before a later run};
\node[f7sc] at (18.82,-2.400) {\S\ref{sec:rq2_dependable}};
\draw[f7rl2] (0.78,-2.630) -- (18.82,-2.630);
\ficon{0.9500000000000001}{-2.860}{f7u}{\icoin}
\node[f7gp] at (1.3,-2.860) {Affordable Operation};
\node[f7as] at (5.1,-2.860) {resource accounting};
\node[f7cg] at (9.4,-2.860) {the resource arrangement};
\node[f7cg] at (13.1,-2.860) {before a run; \textbf{spent inside it}};
\node[f7sc] at (18.82,-2.860) {\S\ref{sec:rq2_affordable}};
\draw[f7rl2] (0.78,-3.090) -- (18.82,-3.090);
\ficon{0.9500000000000001}{-3.320}{f7u}{\ilock}
\node[f7gp] at (1.3,-3.320) {Bounded Reach};
\node[f7as] at (5.1,-3.320) {environment access};
\node[f7cg] at (9.4,-3.320) {the reach};
\node[f7cg] at (13.1,-3.320) {at setup, before any run};
\node[f7sc] at (18.82,-3.320) {\S\ref{sec:rq2_bounded}};
\draw[f7rl2] (0.78,-3.550) -- (18.82,-3.550);
\ficon{0.9500000000000001}{-3.780}{f7u}{\ieye}
\node[f7gp] at (1.3,-3.780) {Reviewability};
\node[f7as] at (5.1,-3.780) {observability, human oversight};
\node[f7cg] at (9.4,-3.780) {the gate, the record};
\node[f7cg] at (13.1,-3.780) {at a run's edges; \textbf{asked inside it}};
\node[f7sc] at (18.82,-3.780) {\S\ref{sec:rq2_reviewability}};
\draw[f7rl2] (0.78,-4.010) -- (18.82,-4.010);
\ficon{0.9500000000000001}{-4.240}{f7u}{\idoor}
\node[f7gp] at (1.3,-4.240) {Equitable Access};
\node[f7as] at (5.1,-4.240) {system access};
\node[f7cg] at (9.4,-4.240) {the install};
\node[f7cg] at (13.1,-4.240) {at install, before the first run};
\node[f7sc] at (18.82,-4.240) {\S\ref{sec:rq2_equitable}};
\draw[f7rl] (0.78,-4.490) -- (18.82,-4.490);
\node[f7lg] at (0.92,-4.720) {In \textcolor{f7acc}{Autonomous Operation} alone, the aspect most discussed and the condition its users changed differ.};
\draw[f7arw] (4.98,-4.94) -- (4.98,-5.74);
\draw[f7dsh] (4.40,-5.66) -- (4.40,-5.3) -- (0.12,-5.3) -- (0.12,-3.780) -- (0.36000000000000004,-3.780);
\node[f7al] at (0.50,-5.140) {a mid-run request\ \ \S\ref{sec:rq2_reviewability}};
\draw[f7bak] (15.60,-5.66) -- (15.60,-5.0200000000000005);
\fill[f7acc,draw=white,line width=1.1pt] (15.60,-5.320) circle (0.21);
\node[f7bn] at (15.60,-5.320) {4};
\node[f7bl] at (15.29,-5.320) {The Change Available};
\fill[white,draw=f7g,line width=0.7pt,rounded corners=6pt] (0.4,-7.05) rectangle (9.55,-5.66);
\fill[f7g,draw=white,line width=1.1pt] (0.8600000000000001,-6.100) circle (0.21);
\node[f7bn] at (0.8600000000000001,-6.100) {2};
\node at (1.3,-6.05) {\includegraphics[width=0.38cm]{openclaw.png}};
\node[f7hd,text=f7g] at (1.58,-6.100) {While the Run Is Under Way};
\node[f7sb] at (0.78,-6.640) {the agent acts; the user is reached only by a request};
\fill[white,draw=f7p,line width=0.7pt,rounded corners=6pt] (10.15,-7.05) rectangle (19.0,-5.66);
\fill[f7p,draw=white,line width=1.1pt] (10.610000000000001,-6.100) circle (0.21);
\node[f7bn] at (10.610000000000001,-6.100) {3};
\ficon{11.05}{-6.100}{f7p}{\ipers}
\node[f7hd,text=f7p] at (11.33,-6.100) {After the Run Finished};
\node[f7sb] at (10.530000000000001,-6.640) {the bill, the record, and verification};
\draw[f7arw] (9.64,-6.355) -- (10.06,-6.355);
\end{tikzpicture}}
\caption{Value-sensitive delegation across the six value groups. Each row gives a group's most discussed agent aspect (Section~\ref{sec:rq1_aspects}), the condition its users changed, and when posts described them changing it. The third and fourth columns are synthesized across Sections~\ref{sec:rq2_autonomous}--\ref{sec:rq2_equitable} rather than paired within user posts. Step~4 marks the return path, in which a user changes a condition for a later run rather than for the run that raised the value.}
\Description{A diagram in four numbered steps. Step one is a table with one row per value group. Each row gives the most discussed agent aspect for that group, the condition users acted on, and when posts described them acting on it. Autonomous Operation most discussed the model core but acted on the reach, widening it before a later run; this row is marked in a contrasting color as the one divergence. Dependable Operation most discussed the model core and acted on the model and host, after a failure and before a later run. Affordable Operation most discussed resource accounting and acted on the resource arrangement, set before a run but spent inside it. Bounded Reach most discussed environment access and acted on the reach, settled at setup before any run. Reviewability most discussed observability and human oversight and acted on the gate and the record, set and read at a run's edges but asked about inside it. Equitable Access most discussed system access and acted on the install, at install and before the first run. Step two: while the run is under way, the agent acts, and the user is reached only by a request. Step three: after the run finished, the user has the bill, the record, and verification. Step four is an arrow returning from step three to step one, labeled the only change available. A dashed arrow runs from step two back to the Reviewability row, labeled a mid-run request.}
\label{fig:conditions}
\end{figure*}
\subsection{Value-Sensitive Delegation: Locating Values in the Conditions of Agent Use}
\label{sec:disc_delegation}
Value Sensitive Design (VSD) treats human values as interactional, meaning that a value takes shape in the relationship between a technology's properties, the people affected by it, and the context where it is used~\cite{friedman2006vsd}. A delegated AI agent like OpenClaw stretches that relationship over time, because the user configures the agent, the task runs autonomously, and the user checks the outcomes afterward. Section~\ref{sec:rq1_aspects} reports that in four of the six value groups, user posts concentrated on agent aspects the user configures rather than on agent aspects the agent exercises without the user. These included what a run could spend (Affordable Operation), what it could reach (Bounded Reach), when it had to ask (Reviewability), and what had to be installed (Equitable Access). While Dependable Operation and Autonomous Operation frequently discussed the model core, the model core was the most discussed agent aspect corpus-wide and did not uniquely set those two value groups apart.

We define an \emph{operating condition}, or \emph{condition} for short, as the configurable boundary a user establishes around an agent's run, together with the time point at which they can act on it (Figure~\ref{fig:conditions}). Reading the six value groups together, we identify five conditions in these posts, namely the model and host, the resource arrangement, the granted reach, the approval gate and record, and the initial install (Sections~\ref{sec:rq2_autonomous}--\ref{sec:rq2_equitable}). A condition is not a further coded category. An agent aspect names the \emph{place} where a value is at stake, and a condition is \emph{what the user can set} at that place. Most agent aspects carry no condition at all, because the agent exercises them independently. A user can choose which model runs, but not how it reasons, and can set a stopping rule, but not the plan the agent forms to meet it. Where a condition does exist, its timing varies, since some are settled at install while others can be reset before each later run.

We name this pattern \emph{value-sensitive delegation}, the relationship between a human value and the operating condition that carries it, together with the moment at which a user can still change that condition. Two claims follow. First, in four of the six value groups the value attached to those conditions rather than to the finished work (Section~\ref{sec:rq1_aspects}). Second, whether a user could protect a human value depended on \textit{when} its condition could be set. In four value groups, the condition can be fixed before a run starts. In Affordable Operation and Reviewability, by contrast, the condition moves while the run is underway, and users in those two groups reported values at the conditions they had fixed in advance but reported costs at the conditions that moved (Sections~\ref{sec:rq2_affordable} and~\ref{sec:rq2_reviewability}). The relationship runs in both directions, since the conditions a user sets bound what the agent may do, and what the agent does then sends the user back to adjust those conditions for the next run (Step~4 in Figure~\ref{fig:conditions}).

In five of the six value groups, the most discussed agent aspect and the condition users acted on align. However, in Autonomous Operation, users most often discussed the model core, yet they acted on the agent's reach (the files, credentials, and services the agent was permitted to touch). What these users discussed and what they could change therefore came apart (Sections~\ref{sec:rq1_aspects} and~\ref{sec:rq2_autonomous}). AI engineering research decomposes agents into the same modules of memory, planning, and tool actions~\cite{wang2024survey,xi2025rise,sumers2023cognitive}, while security research locates risks across architectural layers~\cite{suwansathit2026security,ying2026uncovering}. Both vocabularies identify a property or risk \textit{inside} the agent. Neither indicates which condition a user would change or when that change could occur, which is what Figure~\ref{fig:conditions} provides.

Because these conditions can be reset between runs, users adjust their delegation strategy without abandoning OpenClaw. Prior work examines user reliance on automation~\cite{parasuraman1997humans}, and when such reliance is appropriate~\cite{lee2004trust}. Recent agent research maps the design space for user control~\cite{cheng2026mapping} and balances autonomy with human oversight~\cite{naik2025earlyadopters}. These accounts do not record what a user changed before the next run. Following failures, Dependable Operation users swapped models or hosts rather than discarding OpenClaw, while Autonomous Operation users widened the agent's reach even without failures (Sections~\ref{sec:rq2_dependable} and~\ref{sec:rq2_autonomous}). Value-sensitive delegation thus refines these experiences of reliance~\cite{parasuraman1997humans,lee2004trust} by separating continued agent use from the conditions users changed before the next run.

The Reviewability value group highlights this timing difference by mixing pre-run checks with mid-run interruptions. Prior work studies how user involvement impacts trust, performance, and cognitive load~\cite{bucinca2021trust,he2025plan,zhou2026checking,tang2026darkpatterns}, and distinguishes interventions that halt an agent from those that steer it~\cite{kim2026interventions}. However, our findings separate three mechanisms often collapsed into ``human oversight'': the gate, the record, and the request. A gate is a boundary set once that holds without active watching. A record supports verification after the run. A request, unlike either, requires the user inside a run that is still going. While requests support human judgment, they can also re-ask about a decision the user already approved, or arrive where the user cannot answer them. Separating these three explains why supervision was reported as both costless and costly within the same value group, since gates and records keep the user's authority in force without occupying their time, whereas a request has to be answered during the run (Sections~\ref{sec:rq2_reviewability} and~\ref{sec:rq2_fulfillment}).

The conditions also explain how users could name a value as threatened even when no harm occurred, as seen in Bounded Reach posts concerning granted access (Section~\ref{sec:rq2_bounded}). Value-sensitive delegation thus contributes a concrete unit of analysis to VSD. VSD identifies a human value between what a designer built and what a user brings, and a delegated run holds the two apart in time. A condition is what carries the user's intent across that interval. The user fixes the condition beforehand and it stays in force while the agent works, so it is the user's contribution that is present when the value is realized. A study of a deployed agent can therefore ask which condition a reported value attached to, and when the user could have changed it. A generic list of values supports neither question, because it records what mattered to someone without recording what they could change. Value-sensitive delegation therefore locates a user's remaining agency in the conditions rather than in the run itself.

\subsection{Gains and Costs Users Attributed to Running OpenClaw Autonomously}
\label{sec:disc_gains_costs}
In Autonomous Operation, the gain users described was time they no longer had to spend rather than a better result (Section~\ref{sec:rq2_autonomous}). Human--AI collaboration measures what a user and an AI system achieve together~\cite{he2025plan} and what taking part costs the user~\cite{bucinca2021trust}, but both measurements require the user to be present during the run. User experience investigation~\cite{hassenzahl2006ux}, like our study, focuses on the experience in a user's own state and surroundings rather than in an AI system's properties. A delegated agent like OpenClaw carries that argument to its limit, because the hours a user gets back are hours spent entirely away from the agent. Recent agent studies have moved the same way, looking past the run to what people manage around it~\cite{cheng2026mapping,naik2025earlyadopters,zhang2026consequences}. Our findings give that move an empirical footing, since the benefit these users reported does not appears in what a run produced. One post in Autonomous Operation ran the other way, describing time as shrinking rather than expanding and asking what a user's own effort is worth once the agent does the work (Section~\ref{sec:rq2_autonomous}). A single post settles nothing, but it marks a limit on the gain, because an account that counts the hours a user no longer spends, without asking what the user does with them, would read that post as a success.

The costs users mentioned similarly materialized outside the finished task before anything went wrong. In Affordable Operation, the charge came from the agent's readiness rather than from a requested task. In Bounded Reach, the risk users described was access they had granted, and in both value groups these were the posts where the value was mostly not met (Sections~\ref{sec:rq2_affordable}, \ref{sec:rq2_bounded}, and~\ref{sec:rq2_fulfillment}). Two lines of work describe this situation, and neither treats a granted permission as a cost the user is already paying. Dependable computing counts such a condition as an active fault only once it produces an error, and treats it as dormant until then~\cite{avizienis2004basic}, while security research on these agents judges a deployment by what an adversary could achieve from it~\cite{su2025survey,suwansathit2026security,ying2026uncovering,wang2026assistant,wang2026your}. The user posts in our study part company with both, because for their authors the granted access was already the cost, written down while the deployment was working as intended. The accounts differ in what counts as harm, an event in the literature and an ongoing state in these user posts.

One cost held its direction everywhere. Supervision workload was mostly not met in all six value groups, and it was the only majority-not-met user outcome whose cells cleared the display threshold everywhere (Section~\ref{sec:rq2_fulfillment}). Its direction therefore did not depend on what a value group cared about or on which condition its users acted. Human workload research has long kept result and cost apart, rating how a task went separately from what it cost the user who performed it~\cite{hart1988nasa}, and agent studies have priced human oversight by evaluating specific designer-chosen interventions, such as checkpoints or interruptions~\cite{bucinca2021trust,he2025plan,zhou2026checking,tang2026darkpatterns}. The charge in these user posts has no such author, since it attaches to letting a run proceed at all and survives across six value groups that share relatively little else. Automation research offers the user one remedy, adjusting how much work they hand over~\cite{lee2004trust,parasuraman1997humans}, but that remedy does not address a charge whose direction persists regardless of what the user delegates. We therefore read supervision workload in these posts as a charge of delegation itself rather than as the price of a particular human oversight mechanism.

What did vary was the gain. Where users described what the agent delivered, how often the value held moved with the value group. In three of the six value groups, a leading user outcome that mostly held sat beside one that mostly did not, most widely in Bounded Reach (Section~\ref{sec:rq2_fulfillment}). Value alignment work argues for human values built for a context~\cite{liscio2022values} and for reaching the people whose values are at stake~\cite{birhane2022participatory,kirk2024prism,sadek2025challenges}. Our findings support that argument for the gains and complicate it for the supervision cost. The gains moved with the value group, so a context-specific method would find them, whereas the supervision cost pointed in one direction in every group, even as its magnitude varied, leaving such a method nothing to catch. Holistic evaluation asks for many measures reported together rather than one headline score~\cite{bommasani2023holistic}, and recent surveys of agent evaluation~\cite{mohammadi2025evaluation,yehudai2025survey} and a human-centered evaluation framework~\cite{chen2025toward} describe a field where task completion dominates while cost and safety draw less measurement. User posts in our study thus point to a sharper rule than reporting more numbers. A measure that changes direction with what users care about and a measure that holds one direction should be reported separately, because averaging them hides the one that never turns.

\subsection{Design Implications}
\label{sec:disc_design}
The design question that follows from Sections~\ref{sec:disc_delegation} and~\ref{sec:disc_gains_costs} is not how to make an agent perform better, but what an AI agent should let a user set, and when. Each implication below takes one of the five conditions and answers that question for it, from the narrowest change to the one that decides whether a user can make any of the others.

\textbf{From spending decided inside a run $\to$ Design a ceiling the agent cannot exceed.} Affordable Operation concentrated on resource accounting, and that agent aspect carries the lowest met rate in the user posts (Sections~\ref{sec:rq1_aspects} and~\ref{sec:rq1_fulfillment}). The resource arrangement is therefore where design should start. One user had already assembled the accounting the agent lacked, an alert threshold and scheduled context compaction, and still could not hold consumption down (Section~\ref{sec:rq2_affordable}). An alert posts spending that has already happened, so the user is told about a decision the run has already made. Human--AI design guidelines ask a system to let a user customize what it monitors and how it behaves~\cite{amershi2019guidelines}, thereby giving the user control. However, a global control that declares what an agent will do cannot stop what a run spends once the run begins. AI agents like OpenClaw should therefore accept a ceiling they must run within, not an alert the user must watch. A bill should also name the trigger behind each charge, so that a charge produced by the agent's readiness reads differently from requested work.

\textbf{From reach that ships open $\to$ Design a narrow default and let users widen it.} Bounded Reach concentrated on environment access, and risk exposure led its user outcomes (Sections~\ref{sec:rq1_aspects} and~\ref{sec:rq2_bounded}). These users documented the reach they had granted before rather than harm they had suffered. One user separated the defects found in the software from a default that shipped open. A default of that kind is not a failure produced by a run, so no record of a run surfaces it. The harnesses that evaluate these agents already narrow reach, since a harness fixes what an agent may touch~\cite{liu2024agentbench,zhou2024webarena,xie2024osworld,jimenez2024swebench,yao2025taubench}. That narrowing stays inside the evaluation and does not reach the person who installs the agent. The narrow default should therefore ship in place, leaving widening as the user's action. An agent should also show what a run may reach at the moment a user grants it, so opening reach becomes a decision. Nothing here shows that a narrower default would have prevented an incident.
 
\textbf{From requests that repeat or go unseen $\to$ Design an approval that persists and reaches the user.} Reviewability was the one group whose posts concentrated on two agent aspects, observability and human oversight (Section~\ref{sec:rq1_aspects}). That pair separates looking from answering. Within the same value group, one leading user outcome mostly held and the other mostly did not (Section~\ref{sec:rq2_fulfillment}). Conditions set before a run and records read afterward asked nothing of the user while the run was underway. A request that arrived mid-run cost something, either because the user had already approved that step or because it appeared somewhere they could not see it (Section~\ref{sec:rq2_reviewability}). Work on agent oversight models how much confirmation a run should carry~\cite{zhou2026checking} and what form an intervention should take~\cite{kim2026interventions}. These user posts point to a different variable. What made a request expensive was not its frequency but whether it reopened a settled decision and whether it landed where the user could answer. An approval should therefore persist within the scope the user granted it, so a run stops re-asking about a step that the user authorized. A request should also travel the channel the agent already uses to reach that user, not a dialog on a machine no one watches.
 
\textbf{From swapping the LLM model as the available repair $\to$ Design a completion test the user writes.} The two largest value groups both discussed the model core more than any other agent aspect, without being set apart by it (Section~\ref{sec:rq1_aspects}). The condition available there is substitution, and these posts record users changing which LLM model ran rather than how it behaved. After a failure, they swapped the model or lowered what they asked of it (Section~\ref{sec:rq2_dependable}). A design that offers a better model therefore offers them another substitution rather than a new kind of control. The fix these value groups needed is not in the model or host but in a condition set before a run, because completion no longer evidenced that the work was done and these users had already taken verification back themselves. Benchmark harnesses already solve that problem, declaring what counts as done before a run and having something other than the agent check it~\cite{jimenez2024swebench,yao2025taubench}. That mechanism transfers to a deployment, with one change. The user writes the test, not a benchmark author, so it becomes a condition set around the agent rather than a fixture inside a harness. OpenClaw should therefore accept a completion test that its user writes before a run. A retry and verification policy should sit alongside it as a second condition set in advance.
 
\textbf{From conditions only a user who got in can set $\to$ Design the install as part of the agent.} Equitable Access concentrated on system access and carried the lowest met rate of the six value groups (Sections~\ref{sec:rq1_aspects} and~\ref{sec:rq1_fulfillment}). What these users credit for a working agent is the path rather than the agent at the end of it (Section~\ref{sec:rq2_equitable}). A documented onboarding command made a provider switch easy, a host provider earned credit the agent did not, and other users weighed the install itself, asking whether someone uncomfortable with a command line could run it at all. Every implication above assumes a user who got far enough to set the condition it names. An agent should therefore ship with the spending ceiling, the narrow reach default, the persistent approval, and the completion test already in place and adjustable without a command line, rather than leaving the user to assemble them.

\section{Limitations and Future Work}
\label{sec:limitations}
Our study is observational and post-level. Because we analyzed Reddit posts from OpenClaw's first few months of public use, the people it records are early adopters whose posts may overrepresent experiences worth writing about. Without stable author identifiers, we cannot separate many posts by one user from a single post by each of many users, so every percentage in this paper describes posts rather than users. Our associations also do not establish causal effects. As detailed in Section~\ref{sec:human_validation}, the LLM-assisted coding showed variable reliability across targets, so estimates for infrequent categories carry the most classification uncertainty. Two user outcomes carry a further limit. Assessing whether a user's trust is appropriately calibrated requires knowing both how much they relied on the agent~\cite{parasuraman1997humans} and whether the agent's actual performance justified that reliance~\cite{lee2004trust}. Because a single outcome label applied to a Reddit post cannot capture the agent's unobserved, ground-truth performance, we draw no claims from trust calibration or recovery behavior in the value-group narratives in Section~\ref{sec:rq2_results}, although their cells appear in Figure~\ref{fig:value_outcome_fulfillment_heatmap} and Table~\ref{tab:fulfillment_outcome}. Finally, the six value groups we constructed for this analysis were not validated as a measurement model. These limits mark out our next studies. The coding schema can be applied to posts about other agents and platforms to test which of the patterns here belong to OpenClaw and which belong to delegation itself. User-level designs, such as interviews or diary studies that follow the same user across runs, could test whether the operating conditions that carried the values in these posts also govern individual experience, and could recover the outcomes that public posts undercount. Our findings also suggest a change to agent evaluation. Recording the conditions an agent ran under alongside its task score would let benchmark results speak to the surface where these users registered value success and failure.

\section{Conclusion}
This study presented an LLM-assisted, VSD-grounded content analysis of 73,093 first-person Reddit posts of OpenClaw use. We identified an autonomy that usually held against a dependability that did not, resource accounting as the agent aspect that least often met the value raised against it, and an oversight cost that stayed low at a run's edges but rose once the agent stopped to ask inside one. Our findings call for locating the human values at stake in agent use within the conditions a user sets around a run, namely what it may spend, what it may reach, and when it must ask, rather than within what a run returns, a relationship we name \emph{value-sensitive delegation}. An agent that finishes the task can still fail the user who handed it over.

\bibliographystyle{ACM-Reference-Format}
\bibliography{main}



\appendix
\section{Corpus Construction and Analytic Samples}
\label{app:corpus-construction}

The analytic samples were constructed as follows:
\begin{enumerate}
    \item The source corpus contained 1,100,308 posts.
    \item Stage~1 retained 187,479 candidate posts; the rest were not first-person, were too thin or unclear, or did not yield a usable classification.
    \item Among the candidates included in the final coding run, Stage~2 retained 73,797 first-person posts with the required coding fields and schema-valid output; the rest were not first-person or were too thin or unclear, lacked required coding fields, or failed output validation.
    \item RQ1 uses the 73,093 posts assigned to a predefined value; the other retained posts had open-coded values without a documented mapping to the predefined taxonomy. RQ2 uses the 44,767 posts within the RQ1 sample that contain an attributable user outcome.
\end{enumerate}

\section{Prompt Design for Value-Centered Coding}
\label{app:prompt-design}

The Stage~2 prompt below also asks for an outcome sentiment label. We coded that field but do not analyze or report it in this paper, and no result in the paper depends on it. The specification below summarizes the substantive Stage~2 instructions used to generate the labels analyzed in this study. It presents the coding task, evidence requirements, category definitions and decision boundaries, and primary output fields. Stage~1 used a screening-only prompt that assigned the same three relevance categories and produced no other labels. The ``HAAI refinements'' in the listing are the nine study-specific values described in Section~\ref{sec:data_analysis}.

\begin{lstlisting}[style=codingprompt,caption={Summary of the Stage~2 value-centered coding instructions.}]
Task:
Analyze one Reddit text unit about OpenClaw use. Assign a relevance category and, when the text supports value-relevant evidence, identify one primary human value, the linked agent aspect, value fulfillment, and any attributable user outcome and sentiment.

Evidence and Gating Rules:
1. Code only what the text states or strongly implies. Do not invent values, agent behavior, or outcomes.
2. A value may be explicit or strongly implied by an evaluation, concern, expectation, barrier, tradeoff, control decision, workaround, failure, benefit, or desired design condition. First-person language, a capability statement, or a product description alone does not establish a value.
3. For each first-person unit, re-read the full text once for implicit value evidence before leaving value_mentioned blank.
4. Select one verbatim phrase, sentence, or short multi-sentence excerpt that best supports the primary value. The value, linked agent aspect, fulfillment status, and optional outcome must all be grounded in that excerpt; do not combine unrelated evidence.
5. Assign one primary label per categorical field, keep the taxonomies conceptually distinct, and leave a field blank when evidence is insufficient.
6. If relevance_category is too_thin_or_unclear or no value is supported, leave all substantive fields blank and set outcome_sentiment to null.
7. Code a user outcome only when the shared excerpt states or strongly implies a concrete consequence for the author. A value concern alone is not an outcome.

Relevance Categories:
- first_person_experience: the author's own use, attempted use, concrete setup, experienced problem, help-seeking about that setup, or directly observed agent behavior.
- secondhand_observation: a substantive evaluation or value concern not grounded in the author's own experience.
- too_thin_or_unclear: insufficient context, a passing mention, generic promotion or question, off-topic text, a deleted or truncated fragment, or no substantive observation.
- Product introductions, promotions, link shares, and lists of possible uses remain too_thin_or_unclear unless they contain a specific evaluation or value concern.

Agent Aspect Taxonomy:
- system_access: installation, onboarding, authentication, provider configuration, subscription state, account readiness, billing access required before use, or workspace setup.
- model_core: the underlying LLM, including model choice, provider, version, tier, decoding parameters, backend, or runtime model.
- agent_profile: the system prompt, persona, role definition, identity, or framing instructions that shape agent behavior.
- task_specification: the task goal, scope, boundary, constraint, authority level, approval rule, stopping condition, or success criterion.
- reasoning: visible or user-reported reasoning, reflection, self-critique, deliberation, decision rationale, or intermediate reasoning before acting. Do not infer hidden reasoning.
- planning: task decomposition, goal sequencing, plan formation, or plan revision.
- memory: context-window management, short-term task state, long-term memory, retrieval, summarization, persistence, forgetting, or conversation history.
- tool_selection: the choice of which tool, function, command, or action to invoke.
- tool_execution: invocation of tools, function calls, API calls, browser actions, terminal commands, app actions, or external operations.
- action_effects: downstream changes to user-visible artifacts, including creation, editing, deletion, movement, or modification of files, documents, code, repositories, or data.
- environment_access: sandbox boundaries, file permissions, credential access, local or cloud resources, data-access scope, external accounts, or workspace permissions.
- observability: traces, logs, action history, current-state surfacing, decision rationale, explanation, or progress visibility.
- error_handling: error detection, reporting, debugging, retry, rollback, repair, or recovery after a failed action.
- human_oversight: approval gates, confirmation prompts, intervention points, supervision affordances, or human-in-the-loop controls.
- multi_agent_orchestration: coordination among agents or subagents, including role assignment, agent-to-agent communication, and delegated workers.
- resource_accounting: tokens, compute, quota, rate limits, budget pressure, billing impact, or resource use during or after operation.
- runtime_performance: latency, speed, responsiveness, throughput, stability, or performance under task load.
- other: a substantive agent-design concern outside the taxonomy when no listed aspect fits.

Agent Aspect Boundary Rules:
- system_access concerns getting the system ready for use; environment_access concerns what resources the agent may reach after access is configured.
- tool_execution concerns invoking an operation; action_effects concerns the user-visible change produced by that operation.
- A failed task is not automatically error_handling. Use error_handling only when detection, reporting, retry, rollback, repair, or recovery is central.
- resource_accounting concerns measured or constrained resource use; runtime_performance concerns latency, responsiveness, throughput, or stability.
- Use task_specification only when the excerpt concerns a task goal, scope, constraint, delegated authority, approval rule, stopping condition, or success criterion.
- Select only the component linked to the dominant value-centered excerpt, even when a product description mentions several components.

Human Value Taxonomy:
The initial VSD values are sensitizing concepts. The human-autonomous-agent interaction (HAAI) refinements extend them for recurring conditions in interaction with autonomous agents. Both sets describe stakeholder-relevant conditions rather than agent components, safeguards, or implementation mechanisms.

Initial VSD Values:
- human welfare: the physical, material, and psychological well-being of people directly or indirectly affected by the technology.
- property ownership: people's rights to possess, use, manage, benefit from, transfer, or dispose of information and other assets.
- privacy: a person's claim or entitlement to determine how information about them is accessed, collected, used, and communicated.
- freedom from bias: protection from systematic unfairness produced by pre-existing, technical, or emergent bias.
- universal usability: the ability of people with diverse skills, abilities, resources, platforms, and contexts to use the technology successfully.
- trust: a person's willingness to rely on an agent or its responsible operators while accepting vulnerability to their actions, failures, or betrayal.
- autonomy: a person's ability to decide, plan, and act in pursuit of their own goals.
- informed consent: voluntary agreement made after adequate disclosure and comprehension, with competence and a genuine ability to choose.
- accountability: the state in which actions and outcomes are traceable to people or institutions that can explain and answer for them.
- identity: a person's understanding of who they are over time, including continuity, authorship, expertise, and social or professional role.
- calmness: a peaceful and composed state that is not disrupted by unnecessary interruption, anxiety, or cognitive overload.
- environmental sustainability: the preservation of ecosystems and resources for present needs without compromising future generations.

HAAI Value Refinements:
- meaningful human control: the state in which an agent remains responsive to relevant human reasons and its actions and outcomes remain traceable to people who understand and can assume responsibility.
- transparency: the availability of relevant and intelligible information about an agent's goals, capabilities, limits, state, reasoning, actions, and outcomes.
- repairability: the ability to diagnose, reverse, correct, or recover from errors and unwanted changes.
- contextual integrity: the preservation of context-specific norms governing who sends or receives what information, for which purpose, and under what conditions.
- data sovereignty: the legitimate authority of individuals or collectives over data access, storage, processing, transfer, retention, and deletion.
- affordability: the state in which monetary costs do not unreasonably exclude or burden intended users.
- resource stewardship: the responsible and proportionate use of tokens, compute, quota, energy, network capacity, and related service resources.
- dependability: a justified expectation that the agent will provide correct and consistent service under stated conditions, including reliability, availability, integrity, safety, and maintainability.
- security: protection from unauthorized access, disclosure, manipulation, disruption, or control.

Value Source:
Value source records which part of the taxonomy supplied the primary label; it is not a separate value.
- existing_vsd: value_mentioned matches one of the twelve initial VSD labels.
- haai_refinement: value_mentioned matches one of the nine HAAI refinement labels.
- open_code: neither list represents the human value. Do not open-code agent components, implementation mechanisms, safeguards, emotions, task outcomes, workload, debugging actions, or recovery actions as values.

Value Boundary Rules:
- accountability concerns who must explain or answer for actions; dependability concerns whether the agent works correctly and consistently.
- autonomy concerns a person's capacity to pursue their own goals; meaningful human control concerns governance of delegated agent action.
- data sovereignty concerns authority over data boundaries and lifecycle; resource stewardship concerns proportionate use of computational or service resources.
- transparency concerns intelligible information and visibility; accountability concerns answerability and responsibility.
- affordability concerns monetary access or burden; resource stewardship concerns consumption of computational or service resources.
- trust concerns willingness to rely while vulnerable; dependability concerns functional correctness and consistency.
- privacy concerns a person's claim over information about them; contextual integrity concerns appropriate information flow within a social context; data sovereignty concerns authority over data access and lifecycle.

Value Fulfillment:
- met: the value is supported, enabled, protected, respected, or fulfilled.
- not_met: the value is undermined, violated, threatened, frustrated, absent, requested, or expected but unavailable.
- blank: the excerpt identifies a value but does not establish whether it is met.
- Do not assign fulfillment when value_mentioned is blank.

User Outcome Taxonomy:
- task_effectiveness: success, failure, output quality, correctness, completion, or partial completion.
- time_efficiency: time saved or wasted, faster completion, delay, or interruption.
- resource_burden: cost, tokens, compute, quota, rate-limit pressure, or billing burden experienced as a consequence.
- supervision_workload: monitoring effort, cognitive load, checking, babysitting, or context-management burden.
- affective_response: frustration, confusion, anxiety, calmness, delight, annoyance, or relief.
- trust_calibration: confidence, trust repair, distrust, appropriate reliance, overreliance, or reduced reliance.
- adoption_behavior: continued use, adoption intention, abandonment, disuse, churn, or tool switching.
- risk_exposure: experienced, observed, or clearly anticipated data loss, unwanted change, security exposure, privacy exposure, or compliance consequence.
- recovery_behavior: debugging, rollback, repair, prompt workaround, workflow workaround, or recovery-strategy switching.
- Outcome labels describe consequences, not values. A value concern alone is insufficient for risk_exposure.

Outcome Sentiment:
- -1: negative.
- 0: neutral.
- 1: positive.
- null: no coded outcome; missing is not the same as neutral.

Selected Contrastive Examples:
- Beginner onboarding: "I am an absolute beginner and do not know where to start" can support universal usability:not_met and system_access; leave the outcome blank unless the excerpt states a consequence.
- First-person activity without value evidence: "I am developing custom skills for clients" is first_person_experience, but the statement alone does not establish a value.
- Tool versus effect: "The browser call never runs" supports tool_execution; "the agent deleted my file" supports action_effects.
- Dependability and outcome: repeated incorrect results support dependability:not_met and may support task_effectiveness with negative sentiment.

Output:
Return one JSON object with the core evidence and primary coding fields below.
{
  "relevance_category": "<first_person_experience | secondhand_observation | too_thin_or_unclear>",
  "coding_quote_excerpt": "<verbatim value-centered excerpt, or empty string>",
  "agent_aspect": "<one taxonomy label, or empty string>",
  "value_mentioned": "<one primary value, or empty string>",
  "value_source": "<existing_vsd | haai_refinement | open_code | empty string>",
  "value_fulfillment": "<met | not_met | empty string>",
  "user_outcome": "<one outcome-family label, or empty string>",
  "outcome_sentiment": <-1 | 0 | 1 | null>
}

\end{lstlisting}

\section{Coding Schema Reference}
\label{app:schema}

Table~\ref{tab:value-schema} lists the 21 human values with their value groups, and Table~\ref{tab:outcome-schema} lists the nine user-outcome categories. Values marked with a dagger are the nine study-specific categories developed for this corpus, and their definitions condense the codebook wording used in Stage~2. The remaining 12 come from Friedman et al.'s heuristic list of values with ethical import and carry that list's own definitions~\cite{friedman2006vsd}, with the leading \emph{refers to} dropped and a parenthesis added for how each value applies to a delegated agent. The conceptual boundaries of the study-specific categories follow prior accounts of meaningful human control~\cite{santonidesio2018meaningful}, contextual integrity~\cite{nissenbaum2004privacy}, data sovereignty~\cite{hummel2021data}, and dependable and secure computing~\cite{avizienis2004basic}. Example excerpts are drawn verbatim from the 400-post codebook-development sample described in Section~\ref{sec:human_validation}, except for calmness, environmental sustainability, identity, accountability, informed consent, and freedom from bias, which that sample contained no agreed instance of; those six come from the coded corpus. All were selected for brevity and clarity after review for identifying details.

\begin{table*}[htbp]
\centering
\scriptsize
\caption{The nine user-outcome categories, defined as in Table~\ref{tab:groups}. Each names what a reported consequence concerned; value fulfillment carries its direction. Example excerpts come from the 400-post codebook-development sample.}
\label{tab:outcome-schema}
\begin{tabular}{@{}p{2.1cm}p{4.6cm}p{7.6cm}@{}}
\toprule
\textbf{User outcome} & \textbf{Definition} & \textbf{Example post excerpt} \\
\midrule
Task effectiveness & Whether the agent completed the task and how correct or usable the result was. & ``\textit{I've been using my agent to build a temperature trading bot with mixed results.}'' \\
Time efficiency & Whether the agent saved the user time or cost them time. & ``\textit{Im using an old surface pro 6 and it pretty slow with OpenClaw}'' \\
Resource burden & What a run cost the user in money and metered resources. & ``\textit{OpenClaw made openAI and Anthropic \$100K from my credit card.}'' \\
Supervision workload & The effort of watching, checking, and managing what the agent did. & ``\textit{I prefer cowork but it's too hard to keep a remote instance running smoothly for me.}'' \\
Affective response & How the user felt about the experience. & ``\textit{It's a total mess at my side as everyone is using everything.}'' \\
Trust calibration & How far the user relied on the agent, and whether that reliance was warranted. & ``\textit{If an AI can read your emails, track your behavior, and make decisions for you... At what point do you stop being the one in control?}'' \\
Adoption behavior & Whether the user continued with the agent, intended to, or moved away from it. & ``\textit{I am loving openclaw and I plan to do much more.}'' \\
Risk exposure & Harm to the user's data, systems, or compliance position that a post reported or clearly anticipated. & ``\textit{Just don't give it access to anything important.}'' \\
Recovery behavior & What the user did to repair or work around what the agent did. & ``\textit{I stopped updating open claw because every new version critically broke something and ended up costing me several hours just to get it back on track}'' \\
\bottomrule
\end{tabular}
\end{table*}

\begin{table*}[htbp]
\centering
\scriptsize
\caption{The 21 human values and their six value groups; \textdagger{} marks the nine study-specific categories. Example excerpts come from the 400-post codebook-development sample.}
\label{tab:value-schema}
\begin{tabular}{@{}>{\raggedright\arraybackslash}p{1.4cm}>{\raggedright\arraybackslash}p{1.9cm}p{5.4cm}p{5.3cm}@{}}
\toprule
\textbf{Value group} & \textbf{Value} & \textbf{Definition} & \textbf{Example post excerpt} \\
\midrule
Dependable Operation & Dependability\textdagger & Justified expectation that the agent delivers its intended service correctly and consistently under stated conditions. & ``\textit{OpenClaw feels dumber than Claude... even though it uses Claude?}'' \\
 & Trust & Expectations that exist between people who can experience good will, extend good will toward others, feel vulnerable, and experience betrayal~\cite{friedman2006vsd} (agent: the user's reliance on the agent and its operators, and the vulnerability that reliance creates). & ``\textit{Trust grows once people see how consistently it performs.}'' \\
 & Repairability\textdagger & Ability to diagnose, reverse, correct, and recover from agent errors or unwanted changes. & ``\textit{I had this issue after an openclaw update got fixed with another update}'' \\
\midrule
Autonomous Operation & Autonomy & People's ability to decide, plan, and act in ways that they believe will help them to achieve their goals~\cite{friedman2006vsd} (agent: the goals a user pursues by delegating a run rather than performing the work). & ``\textit{I want my Mac back!}'' \\
 & Human welfare & People's physical, material, and psychological well-being~\cite{friedman2006vsd} (agent: the well-being of the user who delegates work to the agent). & ``\textit{switched from self-hosting openclaw to managed hosting and my weekends came back}'' \\
 & Calmness & A peaceful and composed psychological state~\cite{friedman2006vsd} (agent: a state the agent's interruptions during and between runs do not disturb). & ``\textit{OpenClaw always sends me a Telegram message after each run, even when the script finds nothing new.}'' \\
\midrule
Affordable Operation & Affordability\textdagger & Monetary costs of obtaining and using the agent do not unreasonably exclude or burden users. & ``\textit{I have literally paid money that I cannot use}'' \\
 & Resource stewardship\textdagger & Responsible and proportionate use of finite resources such as tokens, compute, quota, and energy. & ``\textit{I'm trying to figure out the smartest way to assign models based on what the task actually needs.}'' \\
 & Environmental sustainability & Sustaining ecosystems such that they meet the needs of the present without compromising future generations~\cite{friedman2006vsd} (agent: the energy and water the infrastructure a run depends on consumes). & ``\textit{AI data centers consuming offensive amounts of electricity and water is a driving factor in this for me.}'' \\
\midrule
Bounded Reach & Security\textdagger & Protection of people, systems, data, and credentials against unauthorized access, manipulation, or control. & ``\textit{Giving an agent system-level access is a security minefield.}'' \\
 & Privacy & A claim, an entitlement, or a right of an individual to determine what information about himself or herself can be communicated to others~\cite{friedman2006vsd} (agent: what the agent may read about its user and pass on). & ``\textit{OpenClaw is the worst privacy nightmare.}'' \\
 & Identity & People's understanding of who they are over time, embracing both continuity and discontinuity over time~\cite{friedman2006vsd} (agent: continuity of authorship and role once work is delegated). & ``\textit{I'm a professional writer and have been resisting the temptation of using AI to tweak my work.}'' \\
 & Property ownership & A right to possess an object (or information), use it, manage it, derive income from it, and bequeath it~\cite{friedman2006vsd} (agent: the assets and outputs a run consumes or produces). & ``\textit{I'm paying for X amount of tokens, I should be able to burn X amount of tokens via whatever appropriate means I can.}'' \\
 & Data sovereignty\textdagger & Legitimate authority to determine how one's data are accessed, stored, processed, transferred, and deleted. & ``\textit{Local-first control (your data stays yours)}'' \\
 & Contextual integrity\textdagger & Preservation of context-specific norms governing who may send or receive what information. & ``\textit{it keeps memory separated by project, so different topics don't get mixed together}'' \\
\midrule
Reviewability & Transparency\textdagger & Availability of intelligible information about the agent's goals, state, reasoning, actions, and outcomes. & ``\textit{probably staring at logs quite a lot if issues pop up.}'' \\
 & Meaningful human control\textdagger & The agent remains responsive to human reasons, and its outcomes remain traceable to responsible people. & ``\textit{risky actions require human approval}'' \\
 & Accountability & The properties that ensure that the actions of a person, people, or institution may be traced uniquely to the person, people, or institution~\cite{friedman2006vsd} (agent: tracing a run's actions to the person answerable for them). & ``\textit{the llm doesn't make a choice and I effectively need to be the final decision maker AKA fall guy if something bad were to occur.}'' \\
 & Informed consent & Garnering people's agreement, encompassing criteria of disclosure and comprehension (for \emph{informed}) and voluntariness, competence, and agreement (for \emph{consent})~\cite{friedman2006vsd} (agent: agreement to what the agent does and to what it does with a user's data). & ``\textit{why were the AI features ON by default, so opt-out, instead of OFF by default, so opt-in?}'' \\
\midrule
Equitable Access & Universal usability & Making all people successful users of information technology~\cite{friedman2006vsd} (agent: reaching a working agent setup regardless of skill, platform, or resources). & ``\textit{OpenClaw orchestration should be a tool at everyone's disposal.}'' \\
 & Freedom from bias & Freedom from systematic unfairness perpetrated on individuals or groups, including pre-existing social bias, technical bias, and emergent social bias~\cite{friedman2006vsd} (agent: bias in what the agent's underlying model produces or filters). & ``\textit{All the AI sourcing tools we've tried spit out the same people.}'' \\
\bottomrule
\end{tabular}
\end{table*}

\section{Thematic Analysis Codebook}
\label{app:codebook}

Table~\ref{tab:codebook} gives the codebook produced by the thematic analysis described in Section~\ref{sec:statistical_analysis}. Coding was bottom-up within the dimensions the Stage~2 coding had already established, so each value group was read through its two most frequent user outcomes. One researcher labeled the sampled excerpts and grouped the labels into sub-themes, discussing the developing codebook with the research team throughout and revising labels and boundaries after each discussion. The two sub-themes above and the two sub-themes below each rule within a value group form that group's two themes, which Sections~\ref{sec:rq2_autonomous}--\ref{sec:rq2_equitable} report as the bolded claims. The third column gives one excerpt from the sampled posts for each sub-theme, quoted as the poster wrote it.

\begin{table*}[t]
\centering
\scriptsize
\caption{Thematic analysis codebook. Each row pairs a sub-theme with one representative excerpt from the sampled posts. The two sub-themes on either side of a rule within a value group make up one theme, and the twelve themes are the bolded claims in Sections~\ref{sec:rq2_autonomous}--\ref{sec:rq2_equitable}.}
\label{tab:codebook}
\begin{tabular}{@{}>{\raggedright\arraybackslash}p{1.7cm}>{\raggedright\arraybackslash}p{3.3cm}>{\raggedright\arraybackslash}p{8.7cm}@{}}
\toprule
\textbf{Value group} & \textbf{Sub-theme} & \textbf{Representative quotation} \\
\midrule
Autonomous Operation & Extending reach to new surfaces & ``\textit{The biggest unlock was letting the agent improve its own environment.}'' \\
 & Delegating whole workflows rather than steps & ``\textit{So this morning I got OpenClaw to build an automated script that pays for parking at the times I used to get tickets.}'' \\
\cmidrule(l){2-3}
 & Hours returned rather than quality gained & ``\textit{My micro SaaS ops went from 8 hours a week to 45 minutes. Run Lobster (OpenClaw) runs the rest.}'' \\
 & Returned time questioned & ``\textit{Good point. When using n8n, the incremental value of openclaw and the time sink just wasn't worth it.}'' \\
\midrule
Dependable Operation & Silent or partial failure & ``\textit{OpenClaw sees empty content and silently falls through to the next model in your fallback chain --- no error, just the wrong model answering.}'' \\
 & Verification taken back by the user & ``\textit{`Verify before asserting' exists because the agent told me a feature was working when it had never been tested.}'' \\
\cmidrule(l){2-3}
 & Substitution as the available repair & ``\textit{Chat performance of GLM was really good, but I did not get tools to work stably with Ollama. That is why I switched to Qwen. But the performance drop is huge.}'' \\
 & Lowered expectations rather than exit & ``\textit{GPT-oss 20b was kinda useless i found, but I still use it for direct question / answer stuff rather than reasoning.}'' \\
\midrule
Affordable Operation & Consumption outside a requested task & ``\textit{Mine just ate 2.88M tokens in half an hour for simple task like playing a song in youtube, volume change and search for Gemini API key usage.}'' \\
 & Spending visible only after the fact & ``\textit{Been running openclaw agents for a while and kept hitting the same wall. No visibility into cost until the bill arrived.}'' \\
\cmidrule(l){2-3}
 & Re-arranging where the money goes & ``\textit{I was getting tired of how fast OpenClaw burns through money with API tokens, so I switched to running everything locally with Ollama.}'' \\
 & Cost measured against what it displaced & ``\textit{I replaced my \$3900/year sales stack using Claude Code and OpenClaw in 4 days. It now costs me \$40/mo to run.}'' \\
\midrule
Bounded Reach & Granted access as the exposure & ``\textit{The OpenClaw config file was world-readable (644 permissions). Anyone else on the system could see tokens and settings.}'' \\
 & Anticipated rather than experienced harm & ``\textit{I've been running OpenClaw for a few months and I'm increasingly worried about prompt injection. Content filtering didn't work --- is anyone else thinking about this?}'' \\
\cmidrule(l){2-3}
 & Placement and isolation set at setup & ``\textit{That's exactly what I do and it works really well. I have proxmox on my homelab, and create a Linux VM for each agent so there's clear separation}'' \\
 & Boundaries that made a run usable & ``\textit{Because of the precautions I put in place, the bot refused, checked the Telegram ID, and shut the conversation down.}'' \\
\midrule
Reviewability & Control set in advance & ``\textit{Biggest win was having a simple `supervisor' agent that enforces a checklist (sources, confidence, next actions) and a hard stop when tools fail.}'' \\
 & Reading the record afterward & ``\textit{Now I can chat with either agent, check memory, browse skills, and inspect tool calls from the same app.}'' \\
\cmidrule(l){2-3}
 & Requests that reopen settled decisions & ``\textit{It is funny that you think the job is done, but when you open your claude app you found a random approval stall everything and it is already in your allowed list. Just straight frustration.}'' \\
 & Requests that arrive out of reach & ``\textit{when I asking OpenClaw to do stuff, it starts and then suddenly stops giving feedback, I have no idea if it's done, or there were issues.}'' \\
\midrule
Equitable Access & Working setup as the precondition & ``\textit{I've been running OpenClaw as my personal AI assistant that works for my whole immediate family.}'' \\
 & Credit given to the path in & ``\textit{Had my openclaw set up and running within the day and im positive I probably would have given up halfway through if I didnt have the guide.}'' \\
\cmidrule(l){2-3}
 & Install as the decision point & ``\textit{failed at installing openclaw 3 times. gave up and moved to Run Lobster (OpenClaw).}'' \\
 & Setup ability as the gate on who benefits & ``\textit{My problem wasn't OpenClaw itself. It was my co-founder / non-technical members of my team who couldn't deploy without a lot of hand holding.}'' \\
\bottomrule
\end{tabular}
\end{table*}

\section{Human Validation}
\label{app:human-validation}

The separate 50-post Stage~1 relevance-screening pilot reported precision of .900, recall of .783, and $F_1=.837$. Because the pilot was not a corpus-random audit of excluded posts, these metrics do not establish full-corpus recall.

The 400-post codebook-development sample supports two comparisons. A coder decision retained the initial LLM label when the review cell was blank and used the coder-entered replacement when the cell was nonblank. Human--human reliability compares the two coder decisions using exact agreement and pooled Cohen's $\kappa$. LLM--human correspondence treats the initial LLM label as the prediction and each eligible coder decision as a separate reference. For targets other than relevance, the human--human metrics use the 328 posts for which neither coder's relevance decision was too thin or unclear. The LLM--human metrics use 671 individual coder decisions for which that coder's relevance decision was not too thin or unclear. Accuracy, macro-precision, macro-recall, macro-$F_1$, and weighted-$F_1$ therefore compare the LLM labels with coder decisions rather than measure accuracy against an objective gold standard. Metric calculations trimmed outer whitespace but otherwise treated each recorded replacement string as a distinct label; no post hoc class normalization was applied. Table~\ref{tab:human-validation} reports both comparisons.

\begin{table*}[htbp]
\centering
\small
\caption{Human validation in the 400-post codebook-development sample. Blank review cells retain the initial LLM label and nonblank cells use the coder's replacement. Human--human metrics compare the two coder decisions; LLM--human metrics treat coder decisions as references and initial LLM labels as predictions. Macro-P and Macro-R are macro-averaged precision and recall.}
\label{tab:human-validation}

\textit{Panel A: Human--human exact agreement and pooled Cohen's $\kappa$}\\[2pt]
\begin{tabular}{@{}lrrr@{}}
\toprule
\textbf{Coding target} & \textbf{$n$ posts} & \textbf{Exact agreement} & \textbf{$\kappa$} \\
\midrule
Relevance & 400 & 94.75\% & .906 \\
Agent aspect & 328 & 82.01\% & .797 \\
Human value & 328 & 90.85\% & .897 \\
Value fulfillment & 328 & 98.78\% & .979 \\
User outcome & 328 & 92.99\% & .918 \\
\bottomrule
\end{tabular}

\vspace{6pt}
\textit{Panel B: Initial LLM label versus coder decisions}\\[2pt]
\begin{tabular}{@{}lrrrrrr@{}}
\toprule
\textbf{Coding target} & \textbf{$n$ decisions} & \textbf{Accuracy} & \textbf{Macro-P} & \textbf{Macro-R} & \textbf{Macro-$F_1$} & \textbf{Weighted-$F_1$} \\
\midrule
Relevance & 800 & .948 & .948 & .911 & .924 & .946 \\
Agent aspect & 671 & .900 & .895 & .870 & .863 & .894 \\
Human value & 671 & .943 & .622 & .661 & .635 & .937 \\
Value fulfillment & 671 & .994 & .747 & .746 & .747 & .993 \\
User outcome & 671 & .961 & .599 & .609 & .604 & .955 \\
\bottomrule
\end{tabular}
\end{table*}

Coders viewed the initial LLM labels before entering their decisions, so these results measure assisted human--human agreement and agreement between the LLM and coders rather than blind independent coding. The sample was used for codebook development and does not estimate label error across the full corpus.
LLM--human correspondence varied across coding targets: macro-$F_1$ was .635 for human value and .604 for user outcome despite higher accuracy and weighted-$F_1$. We therefore interpret estimates for infrequent categories and cells cautiously; the exact-string calculation treats free-text replacement variants as distinct labels.

\section{Analysis Measures}
\label{app:formal-measures}

Let $O_{ij}$ be the number of posts whose coded label in one field falls in row $i$ and whose coded label in a second field falls in column $j$, with row total $n_{i\cdot}$, column total $n_{\cdot j}$, and grand total $N$. Under independence, the expected count is
\begin{equation}
E_{ij}=\frac{n_{i\cdot}n_{\cdot j}}{N}.
\label{eq:expected-count}
\end{equation}
The Pearson statistic and Cram\'er's $V$ are
\begin{equation}
\chi^2=\sum_i\sum_j\frac{(O_{ij}-E_{ij})^2}{E_{ij}},
\qquad
V=\sqrt{\frac{\chi^2}{N\min(r-1,c-1)}},
\label{eq:chi-cramer}
\end{equation}
where $r$ and $c$ are the table dimensions. Equations~\ref{eq:expected-count} and~\ref{eq:chi-cramer} define the expected counts and post-level association summaries. Every expected count exceeded five in the tables we report, with a minimum of 36.8 in the value-group-by-user-outcome table and 297.6 in the value-fulfillment-by-user-outcome table, so the $\chi^2$ approximation holds in both. The cell-level observed-to-expected (O/E) ratio is
\begin{equation}
L_{ij}=\frac{O_{ij}}{E_{ij}}
=\frac{P(\text{column }j\mid\text{row }i)}{P(\text{column }j)}.
\label{eq:oe-ratio}
\end{equation}
Equation~\ref{eq:oe-ratio} shows that the O/E ratio compares a within-row share with its corresponding corpus-wide share. The within-row percentage $O_{ij}/n_{i\cdot}$, the O/E ratio $L_{ij}$, and the reverse conditional share $O_{ij}/n_{\cdot j}$ answer different questions and are reported separately.

For value fulfillment, let $p_a$ be the corpus-wide met rate at agent aspect $a$, and let $n_{g,a}$ be the number of posts in value group $g$ coded at aspect $a$. The expected number of met posts based on the distribution of agent aspects is
\begin{equation}
E_g^{\mathrm{met}}=\sum_a n_{g,a}p_a.
\label{eq:expected-met}
\end{equation}
With $O_g^{\mathrm{met}}$ denoting the observed number of met posts and $n_g$ the group total, the standardized difference in pp is
\begin{equation}
\Delta_g^{\mathrm{met}}
=\frac{100\left(O_g^{\mathrm{met}}-E_g^{\mathrm{met}}\right)}{n_g}.
\label{eq:delta-met}
\end{equation}
Equations~\ref{eq:expected-met} and~\ref{eq:delta-met} define this expected count and the corresponding observed-minus-expected difference.
A positive value indicates that the observed met rate exceeds the rate expected from the group's distribution of agent aspects; it does not identify a causal group effect.

For fulfillment standardization, we calculated 95\% percentile bootstrap intervals from 2,000 within-group post resamples using seed 7. We held the corpus-wide agent-aspect met rates $p_a$, estimated from the 73,093-post RQ1 sample, fixed across resamples. Thus, the intervals omit uncertainty in those estimates and dependence among posts written by the same author.

For the outcome heatmap, let $n_{g,o}$ denote the number of posts in value group $g$ with user outcome $o$, and let $n_{g,o}^{\mathrm{met}}$ denote the number of those posts whose primary value was coded met. We display the cell met rate as
\begin{equation}
M_{g,o}=100\left(\frac{n_{g,o}^{\mathrm{met}}}{n_{g,o}}\right).
\label{eq:cell-met-rate}
\end{equation}
Equation~\ref{eq:cell-met-rate} defines the heatmap statistic. Values above 50 indicate that met posts outnumber not-met posts within the cell, and values below 50 indicate the reverse. Posts without a coded user outcome do not enter the heatmap, and cells with fewer than 20 posts are marked with an asterisk but are not interpreted.


\end{document}